\documentclass[preprint,5p,times,twocolumn]{elsarticle}

\usepackage{amssymb}
\usepackage{amsthm}
\usepackage{hyperref}
\usepackage{subfigure}
\usepackage[ruled,vlined,linesnumbered]{algorithm2e}
\usepackage{booktabs}
\usepackage{tikz}
\usetikzlibrary{shapes,arrows}
\usepackage{bm}
\usepackage{mathrsfs}
\usepackage{verbatim}
\usepackage{placeins}
\usepackage{overpic}
\usepackage{moreverb,url}

\journal{Parallel Computing}

\begin{document}


\tikzstyle{decision} = [diamond, draw, fill=blue!20, text width=4.5em, 
          text badly centered, node distance=3cm, inner sep=0pt]
\tikzstyle{block} = [rectangle, draw, fill=blue!20, text width=5em, 
          text centered, rounded corners, minimum height=4em]
\tikzstyle{line} = [draw, -latex']
\tikzstyle{cloud} = [draw, ellipse,fill=red!20, node distance=3cm,
minimum height=2em]


\begin{frontmatter}



\title{On the Scalability of CFD Tool for Supersonic Jet Flow 
Configurations}


\author[label1]{Carlos Junqueira-Junior}
\author[label2]{Jo\~{a}o Luiz F. Azevedo}
\author[label3]{Jairo Panetta}
\author[label4]{William R. Wolf}
\author[label5]{Sami Yamouni}
\address[label1]{Arts et M\'{e}tiers Institute of Technology, DynFluid, 
CNAM, HESAM University, 
Paris, France}
\address[label2]{Instituto de Aeron\'{a}utica e Espa\c{c}o, 
S\~{a}o Jos\'{e} dos Campos, Brazil}
\address[label3]{Instituto Tecnol\'{o}gico de Aeron\'{a}utica, 
S\~{a}o Jos\'{e} dos Campos, Brazil}
\address[label4]{Universidade Estadual de Campinas, Campinas, Brazil}
\address[label5]{Data{L}ab Serasa Experian, 
S\~{a}o Paulo, Brazil}


\begin{abstract}
New regulations are imposing noise emissions limitations 
for the aviation industry which are pushing researchers 
and engineers to invest efforts in studying the 
aeroacoustics phenomena. Following this trend, an 
in-house computational fluid dynamics tool is build to 
reproduce high fidelity results of supersonic jet flows 
for aeroacoustic analogy applications. The solver is 
written using the large eddy simulation formulation that 
is discretized using a 
finite difference approach and an explicit time integration. 
Numerical simulations of supersonic jet flows are very 
expensive and demand efficient high-performance computing. 
Therefore, non-blocking message passage interface 
protocols and parallel Input/Output features are 
implemented into the code in order to perform simulations 
which demand up to one billion grid points.
The present work addresses the evaluation of code 
improvements along with the computational performance 
of the solver running on a computer with maximum 
theoretical peak of 2.727 PFlops.
Different mesh configurations, whose size varies from a 
few hundred thousand to approximately one billion grid 
points, are evaluated in the present paper. Calculations 
are performed using different workloads in order to 
assess the strong and weak scalability of the parallel 
computational tool. Moreover, validation results of a 
realistic flow condition are also presented in the 
current work.
	
\end{abstract}


\end{frontmatter}




\newpage

\section{Introduction}
\label{sec:intro}

Exposure to noise can have a huge impact on 
health and induce a range of physiological reactions such 
as an increase in blood pressure, heart rate and breathing. 
Even sudden noise levels commonly experienced in every 
day, such as busy streets, can cause damage to the health 
\cite{Nelson87}. High level of noise emissions on residents 
near airports has forced governments to create laws, 
regulations and guidelines for the certification of 
noise-emitting airplanes \cite{euro-2050,euro-vision2020}. 
According to the new constraints, airplanes which do not 
respect the noise emissions limitations may not be operated 
from all airports or their operators must pay additional fees 
for noise emission. Moreover, noisy aircraft may not be 
operated during the night. Hence, airlines have to consider 
the noise-related airport fees in their operating costs which 
can increase the price of flights. Such a scenario has been 
pushing the civil aviation industry to invest significant 
efforts in studying aeronautical noise emissions. More 
specifically, on the aeroacoustic analogy of supersonic jet 
flows. Such configuration can represent free-jet engine flows 
that contribute significantly to the total sound emissions of 
an airplane \cite{wagner07}.

The authors are interested on the study of unsteady property 
fields of 3-D supersonic jet flow configurations which can 
provide important information in order to 
eventually understand the acoustic phenomena.
Experimental techniques used to evaluate such flow 
configuration are complex and require considerably expensive 
apparatus. Therefore, the authors have developed a numerical 
tool, JAZzY \citep{Junior16}, based on the large eddy simulation 
(LES) formulation \citep{Garnier09} in order to perform 
time-dependent simulations of supersonic compressible jet 
flows. The large eddy simulation approach has been successfully 
used by the scientific community and can provide high fidelity 
numerical data for aeroacoustic applications \citep{Bodony05i8, 
Wolf_doc, lo2012, wolf2012, Junior18-abcm}. The numerical tool 
is written in the Fortran 90 standards coupled with Message Passing 
Interface (MPI) features \citep{Dongarra95}. The HDF5 \citep{folk11,
folk99} and CGNS libraries \citep{cgns2012,legensky02,Poirier00,
Poirier98} are included into the numerical solver in order to 
implement a hierarchical data format (HDF) and to perform 
Input/Output operations efficiently. 

Large eddy simulations of supersonic flows require efficient 
use of significant amount of computational resources in order 
to provide trustworthy results at an acceptable cost. 
Hence, the LES tool is continually improved. The 
latest code release includes modifications in MPI data exchanges 
and reading/writing routines. Classical blocking communicators 
are replaced by asynchronous protocols and the sequential 
mesh reading is re-written in a parallel fashion. Moreover, 
the partitioning routine is optimized regarding the optimal 
workload. 
The present work addresses the computational performance 
evaluation of the in-house numerical tool on a Brazilian 
scientific computer named as Santos Dumont \cite{SDumont}. 
The HPC system was cited on the top 500 list 
\citep{Top500-site,Top500-book} from 2015 to 2017 with a 
maximal LINPACK \citep{Dongarra92,Dongarra88-linpack} 
performance achieved of 456.8 TFlops and a theoretical peak 
performance of 657.5 TFlops. 
A recent upgrade brings the supercomputer up to the 193rd 
position on the top 500 list of November 2019. This new 
version delivers maximal LINPACK and theoretical peak 
performances of 1.849 PFlops and 2.727 PFlops, respectively.
Strong and weak scalability tests are performed in the present 
work using the TFlops version of the cluster. 
Numerical simulations of a supersonic jet flow configuration 
are addressed as a test case in order to evaluate the JAZzY code 
using up to approximately three thousand computational cores 
in parallel for problems with up to one billion 
grid points.

The present article is structured into an introduction followed
by a description of the computer architecture. Then, the numerical 
formulation used by the JAZzY code and the latest
results of supersonic jet flow simulations achieved by the LES 
solver are presented to the reader. In the sequence, a detailed 
discussion about the parallel implementation and features of the 
code is performed followed by the results of the scalability study 
performed in the Santos Dumont supercomputer. Numerical results 
achieved during the validation of the numerical tool are also 
presented here. In the end, the reader can find the concluding 
remarks section and the acknowledgements.



\section{Computer Configuration}

Santos Dumont supercomputer was acquired from Atos HPC systems
by \cite{SDumont} in 2015 by the National Laboratory for 
Scientific Computing (LNCC) \cite{LNCC} in the city of 
Petr\'{o}polis, state of Rio de Janeiro, Brazil. The main 
idea of the laboratory, which is 
affiliated with the Ministry of Science, Technology, 
Innovations and Communications (MCTIC) in Brazil is 
to provide computational resources for research from different 
areas of study such as Engineering, Astronomy, Biology, Chemistry, 
Geosciences and Linguistics.

The full HPC system is in the 193rd position of the top 500 list 
of November 2019 \cite{Top500-site} with a maximal LINPACK 
\citep{Dongarra92,Dongarra88-linpack} performance achieved of 
1.849 PFlops and a theoretical peak performance of 2.727 PFlops.
The scalability of the CFD solver is evaluated using a section 
of the supercomputer with maximal LINPACK performance and 
theoretical peak performance of 456.8 TFlops and 657.5 TFlops, 
respectively. This Teraflop partition presents 18,144 computational 
cores (CPU) spread among 756 nodes (24 computational cores per node).
Graphic processing units (GPU) and Xeon PHI accelerators are also 
coupled to some of the available computing nodes. Moreover, this
partition provides a fat-node with 240 computational cores and 6 TB 
of rapid access memory. A detailed description of computing nodes 
configuration is presented in Tab.\ \ref{tab:sd-computer}. The 
computational resource has a {\it Lustre\textsuperscript{\textregistered}} 
\cite{lustre} file system with a storage capacity of approximately 1.7 
PBytes. There is also a secondary archive system with a storage capacity 
of approximately 640 TBytes. The 756 nodes of the cluster are 
inter-connected by a infiniband network. Red Hat Enterprise 
Linux \cite{redhat} is the operating system of the cluster and 
Slurm \cite{slurm} is used as workload manager.
\begin{table}[htb!]
\small\sf\centering
\caption{Santos Dumont Teraflop partition configuration.}\label{tab:sd-computer}
\begin{tabular}{llll}
\hline
Nodes & Processor & Memory & Nb.\ Cores \\
\hline
504 & \scriptsize 2 x CPU {\it Intel\textsuperscript{\textregistered}}
              {\it Xeon\textsuperscript{\textregistered}}
E5-2695v2 & 64GB & 24 \\ 
198 & \scriptsize 2 x CPU {\it Intel\textsuperscript{\textregistered}}
              {\it Xeon\textsuperscript{\textregistered}}
      E5-2695v2 & 64GB & 24 \\ 
    & \scriptsize  + 2 x GPU {\it Nvidia\textsuperscript{\textregistered}}
      K-40 & 12GB & 5760 \\ 
54  & \scriptsize 2 x CPU {\it Intel\textsuperscript{\textregistered}}
              {\it Xeon\textsuperscript{\textregistered}}
      E5-2695v2 & 64GB & 24 \\ 
    & \scriptsize + 2 x {\it Intel\textsuperscript{\textregistered}}
          {\it Xeon Phi\textsuperscript{\texttrademark}}
      7120 & 16GB & 122 \\ 
1   & \scriptsize 16 x CPU {\it Intel\textsuperscript{\textregistered}}
      Ivy 2.4GHz & 6TB & 240 \\ 
\hline
\end{tabular}
\end{table}




\section{Large Eddy Simulation Formulation}
\label{sec:form}

The numerical simulations of supersonic jet flow configurations are
performed based on the large eddy simulation formulation \cite{Garnier09}. 
This set of equations is based on the principle of scale separation over 
the governing equations used to represent the fluid dynamics, the 
Navier-Stokes formulation. Such scale separation procedure is addressed 
as a filtering procedure in a mathematical formalism. The idea is to 
filter the small turbulent structures and to calculate the bigger ones. 
The Navier-Stokes equations, using the filtering procedure of Vreman 
\cite{Vreman1995}, is written in the current work as
\begin{equation}
\begin{array}{c}
\displaystyle \frac{\partial \overline{\rho} }{\partial t} 
+ \frac{\partial}{\partial x_{j}} 
\left( \overline{\rho} \widetilde{  u_{j} } \right) = 0 \, \mbox{,}\\
\displaystyle \frac{\partial}{\partial t} 
\left( \overline{ \rho } \widetilde{ u_{i} } \right) 
+ \frac{\partial}{\partial x_{j}} 
\left( \overline{ \rho } \widetilde{ u_{i} } \widetilde{ u_{j} } \right)
+ \frac{\partial }{\partial x_{j}} \left( \overline{p} \delta_{ij} \right)
- \frac{\partial {\tau}_{ij}}{\partial x_{j}}  
= 0 \, \mbox{,} \\ 
\displaystyle \frac{\partial \overline{e}}{\partial t} 
+ \frac{\partial}{\partial x_{j}} 
\left[ \left( \overline{e} 
+ \overline{p} \right)\widetilde{u_{j}} \right]
- \frac{\partial}{\partial x_{j}}\left({\tau}_{ij} \widetilde{u_{i}} \right)
+ \frac{\partial {q}_{j}}{\partial x_{j}} = 0 \, \mbox{,}
\end{array}
\label{eq:les}
\end{equation}
in which $t$ and $x_{i}$ are independent variables 
representing time and spatial coordinates of a 
Cartesian coordinate system $\textbf{x}$, respectively. 
The components of the velocity vector $\textbf{u}$ are 
written as $u_{i}$, and $i=1,2,3$. Density, pressure and 
total energy per mass unit are denoted by $\rho$, $p$ and 
$e$, respectively. The $\left( \overline{\cdot} \right)$ and
$\left( \tilde{\cdot} \right)$ operators are used in order 
to represent filtered and Favre averaged properties, 
respectively. The filtered total energy per mass unit 
\cite{Vreman1995} can be written as 
\begin{equation}
	\overline{e} = \frac{\overline{p}}{\gamma - 1} 
	+ \frac{1}{2} \rho \widetilde{u}_{i} \widetilde{u}_{i} \, 
	\mbox{.} 
\end{equation}
The heat flux, $q_{j}$, is written as a function of the static 
temperature, $T$, and the thermal conductivity, $\kappa$,  
\begin{eqnarray}
	{q}_{j} = 
	- \kappa 
	\frac{\partial \widetilde{T}}{\partial x_{j}} 
  & \mbox{where} & \kappa = \frac{\mu C_{p}}{Pr} \, \mbox{.}
  \label{eq:q_mod}
\end{eqnarray}
The thermal conductivity is a function of the specific heat at 
constant pressure, $Cp$, of the Prandtl number, $Pr$, which is 
equal to $0.72$ for air, and of the dynamic viscosity, $\mu$, 
that can be calculated using the Sutherland law,
\begin{equation}
  \mu \left( \widetilde{T} \right) = \mu_{\infty} 
  \left( \frac{\widetilde{T}}{\widetilde{T}_{\infty}}
  \right)^{\frac{3}{2}} 
  \frac{\widetilde{T}_{0}+S_{1}}{\widetilde{T}+S_{1}} 
  \, \mbox{,}
\label{eq:sutherland}
\end{equation}
in which $\mu_{\infty}$, $\widetilde{T}_{\infty}$, 
$\widetilde{T}_{0}$ and $S_{1}$ are reference values. 
Density, static pressure and static temperature are correlated 
by the equation of state given by
\begin{equation}
  \overline{p} = \rho \left(C_{p} - C_{v}\right) \widetilde{T} 
  \, \mbox{,}
\end{equation}
%
where $C_{v}$ is the specific heat at constant volume.
The shear-stress tensor, $\tau_{ij}$, is written 
according to the Stokes hypothesis as
%
\begin{equation}
  {\tau}_{ij} = \mu \,\left( 
	  \frac{\partial \tilde{u}_{i}}{\partial x_{j}} 
	+ \frac{\partial \tilde{u}_{j}}{\partial x_{i}} 
    - \frac{2}{3} \delta_{ij} 
	\frac{\partial \tilde{u}_{k}}{\partial x_{k}}
    \right) \,
  \label{eq:tau_mod}
\end{equation}
%

The large eddy simulation set of equations can be written in 
a more compact form as
\begin{equation}
  \frac{\partial \textbf{Q}}{\partial t} = -\textbf{RHS} \, \mbox{,}
  \label{eq:vec-les}
\end{equation}
where $\textbf{Q}$ stands for the convervative properties vector
and $\textbf{RHS}$ represents the right hand side of 
Eq.\ \ref{eq:les}, given by
\begin{eqnarray}
  \textbf{Q} = \left[ 
  \overline{\rho} \, \mbox{,} \,
  \overline{ \rho } \widetilde{ u_{i} } \,\mbox{,} \, 
  \overline{e}
  \right]^{T} \, & \mbox{and} &
  {RHS}_{i} = 
  \frac{\partial {E}_{i}}{\partial x_{j}} -
  \frac{\partial {F}_{i}}{\partial x_{j}} \, \mbox{.}
  \label{eq:con-prop}
\end{eqnarray}
%
The components of inviscid and viscous flux vectors are
respectively denoted by $E_{i}$ and $F_{i}$, and written as  
\begin{equation}
  \begin{array}{ccc}
    \textbf{E}=\left[
      \begin{array}{c}
        \overline{\rho} \widetilde{u_{j}}\\
  	    \overline{\rho} \widetilde{u_{i}} \widetilde{u_{j}} +
        \overline{p}\delta_{ij} \\
        \left[ \left( \overline{e} + \overline{p} \right)
        \widetilde{u_{j}} \right]
      \end{array}
    \right] & 
    \mbox{and} &
    \textbf{F}=\left[
      \begin{array}{c}
        0 \\
  	    {\tau}_{ij}\\
        {\tau}_{ij} \widetilde{u_{i}} - {q}_{j}
      \end{array}
    \right] \, \mbox{.} 
  \end{array}
\end{equation}

Spatial derivatives are calculated in a structured finite difference 
context and the formulation is re-written for general curvilinear 
coordinate system \cite{Junior16}. The numerical flux is computed 
through a second-order central difference scheme with the explicit 
addition of anisotropic scalar artificial dissipation \cite{Turkel_Vatsa_1994}.
The time marching method is an explicit 5-stage Runge-Kutta 
scheme \cite{jameson_mavriplis_86,Jameson81}.

\begin{figure}[htb]
  \begin{center}
	\subfigure[Lateral view of boundary conditions.]{
    \includegraphics[width=0.45\textwidth,trim=0.25cm 0.25cm 0.25cm 0.25cm,clip]
    {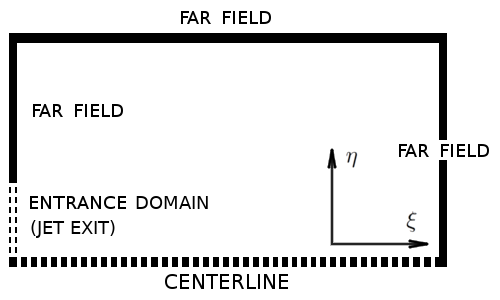} 
    \label{fig:bc-1}
    }
    \subfigure[Frontal view of boundary conditions.]{
    \includegraphics[width=0.45\textwidth,trim=0.25cm 0.25cm 0.25cm 0.25cm,clip]
    {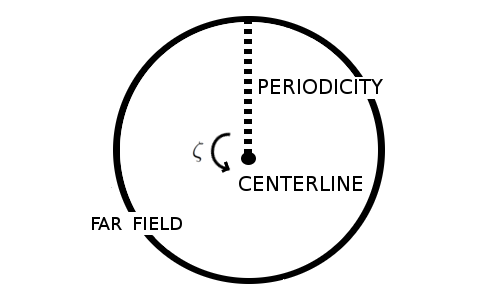} 
    \label{fig:bc-2}
    }
    \caption{2-D Lateral and frontal cuts of the computational domain indicating 
    the position of boundary conditions.}
    \label{fig:bc}
  \end{center}
\end{figure}

Boundary conditions for the LES formulation are imposed in order 
to represent a supersonic jet flow into a 3-D computational domain 
with cylindrical shape. Figures \ref{fig:bc-1} and \ref{fig:bc-2} 
indicate the boundary conditions used in the current work on lateral 
and frontal 2-D cuts of the domain. A supersonic flow 
is implemented at the entrance of the domain, also known by 
the aeroacoustic community as the jet exit of the nozzle, and 
Riemann invariants \cite{Long91} are used to calculate the far field 
boundary conditions.
The centerline of the computational domain is a 
singularity and it requires special treatment. Therefore, conserved 
properties are extrapolated from the adjacent longitudinal plane and 
they are averaged in the azimuthal direction in order to define the updated 
properties at the centerline of the jet.
For the periodic plane, superposed points are used in the 
first and last points in the azimuthal direction in order to close the 
3-D computational since it is a inner surface. The reader can find 
further details about the formulation in the work of {\em Junqueira-Junior
et. al.} \cite{Junior18-abcm}. 




\section{Implementation Aspects}
\label{sec:impl}

The current section presents details of the LES solver and discusses 
the high performance computing implementations introduced into the 
code. JAZzY is developed using the Fortran 90 standard and the 
parallel I/O concept.
Recent improvements of the code include the 
implementation of asynchronous inter-partition communications 
and the development of a preprocessing tool that performs 
an optimized partitioning and creates multiple input grid files 
using a binary tree data structure. Each MPI rank reads its own 
input file when using the new preprocessing routine. 
The I/O modifications are also a first effort to bypass the 
original serial mesh allocation implemented in the code. 


\subsection{Preprocessing Mesh Partitioner}

The LES solver presents a parallel I/O feature in which each MPI 
partition reads its correspondent portion of the mesh. Therefore, a 
3-D grid partitioner is developed in order to provide partitioned 
mesh files to the solver. This preprocessing code can also generate 
grid files for simple geometric configurations. The mesh partitions 
are written using the CGNS standard \citep{cgns2012,legensky02,
Poirier00, Poirier98} which is built on the HDF5 library 
\citep{folk11,folk99}. This library is a general scientific format 
adaptable to virtually any scientific or engineering application. 
It provides tools to efficiently read and write data structured in a 
binary tree fashion. This data structure can handle many types of 
queries very efficiently \citep{Bentley-1975,Bentley-1979}, 
such as time-dependent CFD solutions. 

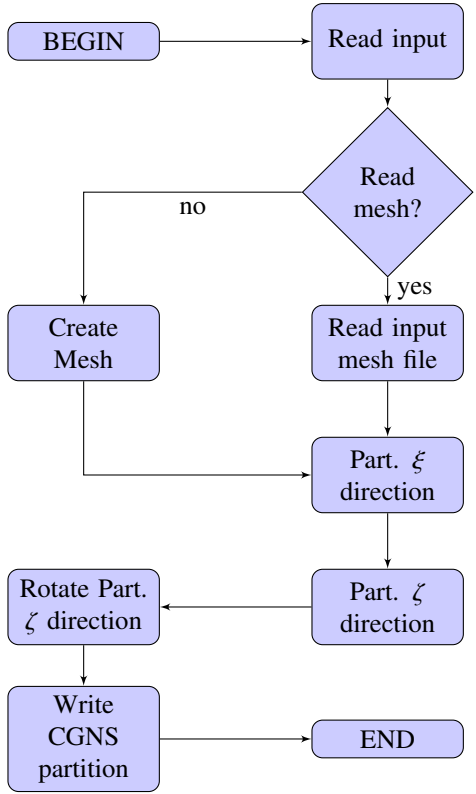
\begin{figure}[htb]
  \begin{center}
    \begin{tikzpicture}[node distance = 1.75cm, auto]
      \node [block,minimum size=0.5cm] (init) {BEGIN};
      \node [block, right of=init,minimum size=1cm,node distance=4cm] 
      (ReadInput) {Read input};
      \node [decision, below of=ReadInput,node distance=2.0cm,minimum size=1cm] 
      (decide1) {Read mesh?};
      \node [block, below of=decide1,node distance=2.0cm,minimum size=1cm] 
      (ReadMesh) {Read input mesh file};
      \node [block, left of=ReadMesh,node distance=4cm,minimum size=1cm] 
      (CreateMesh) {Create Mesh};
      \node [block, below of=ReadMesh,minimum size=1cm] (BalAxi) 
      {Part. $\xi$ direction};
      \node [block, below of=BalAxi,minimum size=1cm] (BalAzi) 
      {Part. $\zeta$ direction};
      \node [block, left of=BalAzi,minimum size=1cm,node distance=4cm] (Rot) 
      {Rotate Part. $\zeta$ direction};
      \node [block, below of=Rot,minimum size=1cm] (WriteCGNS) 
      {Write CGNS partition};
      \node [block, right of=WriteCGNS,minimum size=0.5cm,node distance=4cm] 
      (END) {END};
      \path [line] (init) -- (ReadInput);
      \path [line] (ReadInput) -- (decide1) ;
      \path [line] (decide1) -- node {yes}(ReadMesh);
      \path [line] (decide1) -| node [near start]{no}(CreateMesh);
      \path [line] (ReadMesh) -- (BalAxi) ;
      \path [line] (CreateMesh) |- (BalAxi) ;
      \path [line] (BalAxi) -- (BalAzi) ;
      \path [line] (BalAzi) -- (Rot) ;
      \path [line] (Rot) -- (WriteCGNS) ;
      \path [line] (WriteCGNS) -- (END) ;
    \end{tikzpicture}
  \end{center}
  \caption{Preprocessing flow chart.}\label{fig:preproc}
\end{figure}

Figure \ref{fig:preproc} presents the flow chart of the mesh 
partitioner code. The preprocessing code can read a 2-D mesh file 
from a commercial grid generator or create a 2-D geometric 
configuration along with a grid point distribution using parameters 
provided by the user. Once the mesh is read by the preprocessing 
code, the 2-D domain is partitioned in the axial direction. After 
this procedure, the partitioned mesh is extruded in the azimuthal 
direction, respecting the positioning of the MPI partitions. 
Each portion of the mesh is written using the CGNS standard. 

\begin{figure}[htb]
  \begin{center}
    \subfigure[2-D partitioning in the axial and 
    azimuthal directions.]{
	\includegraphics[width=0.45\textwidth,trim=0.5cm 0.5cm 0.5cm 0.5cm,clip]{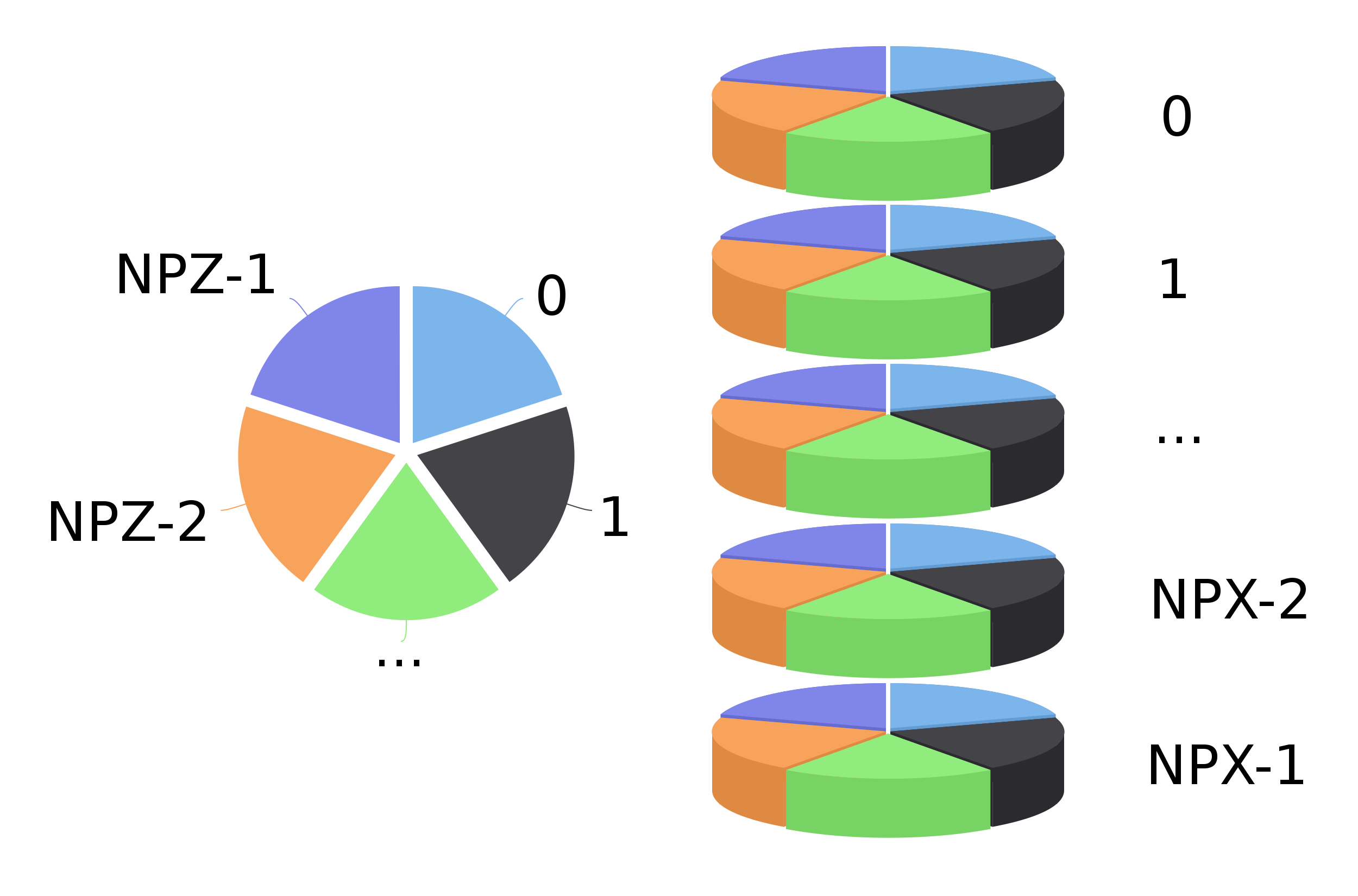}\label{fig:XZ_1}
    }
    \subfigure[Mapping of the 2-D partitioning.]{
      \resizebox{0.45\textwidth}{!}{
      \begin{tikzpicture}[node distance=2.8cm, auto]
        \node [block,minimum size=1.45cm,text width=7em] (i0j0) {0};
        \node [block,minimum size=1.45cm,text width=7em,right of=i0j0] 
		(i0j1) {NPZ};
        \node [block,minimum size=1.45cm,text width=7em,right of=i0j1] 
		(i0j2) {2*NPZ};
        \node [block,minimum size=1.45cm,text width=7em,right of=i0j2] 
		(i0j3) {};
        \node [block,minimum size=1.45cm,text width=7em,right of=i0j3] 
		(i0j4) {};
        \node [block,minimum size=1.45cm,text width=7em,below of=i0j0,
		node distance=1.55cm] (i1j0) {1};
        \node [block,minimum size=1.45cm,text width=7em,right of=i1j0] 
		(i1j1) {NPZ+1};
        \node [block,minimum size=1.45cm,text width=7em,right of=i1j1] 
		(i1j2) {(2*NPZ)+1};
        \node [block,minimum size=1.45cm,text width=7em,right of=i1j2] 
		(i1j3) {};
        \node [block,minimum size=1.45cm,text width=7em,right of=i1j3] 
		(i1j4) {};
        \node [block,minimum size=1.45cm,text width=7em,below of=i1j0,
		node distance=1.55cm] (i2j0) {(...)};
        \node [block,minimum size=1.45cm,text width=7em,right of=i2j0] 
		(i2j1) {(...)};
        \node [block,minimum size=1.45cm,text width=7em,right of=i2j1] 
		(i2j2) {(...)};
        \node [block,minimum size=1.45cm,text width=7em,right of=i2j2] 
		(i2j3) {};
        \node [block,minimum size=1.45cm,text width=7em,right of=i2j3] 
		(i2j4) {};
        \node [block,minimum size=1.45cm,text width=7em,below of=i2j0,
		node distance=1.55cm] (i3j0) {NPZ-2};
        \node [block,minimum size=1.45cm,text width=7em,right of=i3j0] 
		(i3j1) {NPZ+NPZ-2};
        \node [block,minimum size=1.45cm,text width=7em,right of=i3j1] 
		(i3j2) {2*NPZ+NPZ-2};
        \node [block,minimum size=1.45cm,text width=7em,right of=i3j2] 
		(i3j3) {j*(NPZ)+i};
        \node [block,minimum size=1.45cm,text width=7em,right of=i3j3] 
		(i3j4) {};
        \node [block,minimum size=1.45cm,text width=7em,below of=i3j0,
		node distance=1.55cm] (i4j0) {NPZ-1};
        \node [block,minimum size=1.45cm,text width=7em,right of=i4j0] 
		(i4j1) {NPZ+NPZ-1};
        \node [block,minimum size=1.45cm,text width=7em,right of=i4j1] 
		(i4j2) {2*NPZ+NPZ-1};
        \node [block,minimum size=1.45cm,text width=7em,right of=i4j2] 
		(i4j3) {(...)};
        \node [block,minimum size=1.45cm,text width=7em,right of=i4j3] 
		(i4j4) {(NPX*NPZ)-1};
      \end{tikzpicture}\label{fig:XZ_2}
      }
    }
    \caption{2-D partitioning and mapping.}\label{fig:comm_2d}
  \end{center}
\end{figure}

The partitioning of the domain is performed in a matrix fashion in 
order to create structured blocks which can be mapped and easily 
accessed through the use of message protocols. Figure \ref{fig:XZ_1} 
illustrates the segmentation of the domain into the axial and 
azimuthal directions while Fig.\ \ref{fig:XZ_2} presents the mapping 
of the domain. The index of each partition, indicated in 
Fig.\ \ref{fig:XZ_2}, is based on a matrix index system in which the 
rows represent the position in the axial direction and the columns 
represent the position in the azimuthal direction. The partition 
index starts at zero to be consistent with the message passing 
interface standard. NPX and NPZ denote the number of partitions in 
the axial and azimuthal directions, respectively. 

\begin{figure}[htb]
  \begin{center}
    \includegraphics[width=0.35\textwidth,trim=0.1mm 0.1mm 0.1mm 0.1mm,clip]
	{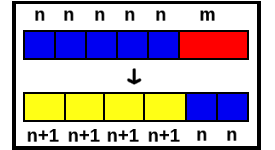}
    \caption{Balancing procedure performed during the
    patitioning of the mesh.}\label{fig:balance}
  \end{center}
\end{figure}

An optimized partitioning is necessary in order to have a well 
balanced distribution of tasks for all resources requested. The 
number of points within a given mesh can be a measure of 
computational costs for CFD applications using the finite difference 
spatial discretization. Therefore, the division of the mesh in the 
axial and azimuthal directions is performed towards a well balanced 
distribution of points. Firstly, the total number of grid points in 
one direction is divided by the number of domains in the same 
direction. The remaining points are spread among the partitions 
when the division is not exact. The same procedure is performed 
in the other directions. Figure \ref{fig:balance} illustrates the 
balancing procedure performed in each direction during the 
partitioning of the computational grid, in which $n$ stands for the 
integer part of the division and $m$ represents the number of points 
in the unbalanced partition.



\subsection{Large Eddy Simulation Code}

\begin{figure}[htb]
  \begin{center}
    \resizebox{0.45\textwidth}{!}{
      \begin{tikzpicture}[node distance = 1.5cm, auto]
        \node [block,minimum size=0.5cm] (init) {BEGIN};
        \node [block, left of=init,minimum size=1cm,node distance=2.5cm] 
		(ReadInput) {Read Input};
        \node [block, left of=ReadInput,minimum size=1cm,node distance=2.5cm] 
		(ReadMesh) {Read Mesh Part.};
        \node [block, below of=ReadMesh,minimum size=1cm] (CalcJacob) 
		{Calc. Jacob. terms};
        \node [block, right of=CalcJacob,minimum size=1cm,node distance=2.5cm] 
		(CalcMetric) {Calc. Metric Terms};
        \node [block, right of=CalcMetric,minimum size=1cm,node distance=2.5cm] 
		(SetInitCond) {Set Init. Cond.};
        \node [block, below of=SetInitCond,minimum size=1cm] (UpdateIt) 
		{Update IT.};
        \node [block, below of=UpdateIt,minimum size=1cm] (UpdateRK) 
		{Update RK.};
        \node [block, below of=UpdateRK,minimum size=1cm] (AsyncCom) 
		{Async. Comm.};
        \node [block, left of=AsyncCom,minimum size=1cm,node distance=2.5cm] 
		(InvFlux) {Inv. Flux};
        \node [block, left of=InvFlux,minimum size=1cm,node distance=2.5cm] 
		(AsyncCom1) {Async. Comm.};
        \node [block, below of=AsyncCom1,minimum size=1cm,node distance=2cm] 
		(Dissip) {Art. Dissip.};
        \node [block, right of=Dissip,minimum size=1cm,node distance=2.5cm] 
		(ViscFlux) {Visc. Flux};
        \node [decision, right of=ViscFlux,minimum size=1cm,node distance=2.5cm] 
		(TestRK) {End RK.};
        \node [block, below of=TestRK,minimum size=1cm,node distance=2cm]
		(UpdateSol) {Update Sol.};
        \node [block, left of=UpdateSol,minimum size=1cm,node distance=2.5cm] 
		(AsyncCom2) {Async. Comm.};
        \node [decision, left of=AsyncCom2,node distance=2.5cm,minimum size=1cm] 
		(TestIt) {End IT.};
        \node [block, below of=TestIt,minimum size=1cm,node distance=2cm] 
		(end) {END};
        \path [line] (init) -- (ReadInput);
        \path [line] (ReadInput) -- (ReadMesh) ;
        \path [line] (ReadMesh) -- (CalcJacob) ;
        \path [line] (CalcJacob) -- (CalcMetric) ;
        \path [line] (CalcMetric) -- (SetInitCond) ;
        \path [line] (SetInitCond) -- (UpdateIt) ;
        \path [line] (UpdateIt) -- (UpdateRK) ;
        \path [line] (UpdateRK) -- (AsyncCom) ;
        \path [line] (AsyncCom) -- (InvFlux) ;
        \path [line] (InvFlux) -- (AsyncCom1) ;
        \path [line] (AsyncCom1) -- (Dissip) ;
        \path [line] (Dissip) -- (ViscFlux) ;
        \path [line] (ViscFlux) -- (TestRK) ;
        \path [line] (TestRK) -- +(+1.65,0) |- node [near start]{no} (UpdateRK) ;
        \path [line] (TestRK) -- node [near start]{yes}(UpdateSol) ;
        \path [line] (UpdateSol) -- (AsyncCom2) ;
        \path [line] (AsyncCom2) -- (TestIt) ;
        \path [line] (TestIt) -- +(-1.65,0) |- (UpdateIt) node [near start]{no} ;
        \path [line] (TestIt) -- node [near start]{yes}(end) ;
      \end{tikzpicture}
    }
  \end{center}
  \caption{JAZzY solver flow chart.}\label{fig:jazzy}
\end{figure}
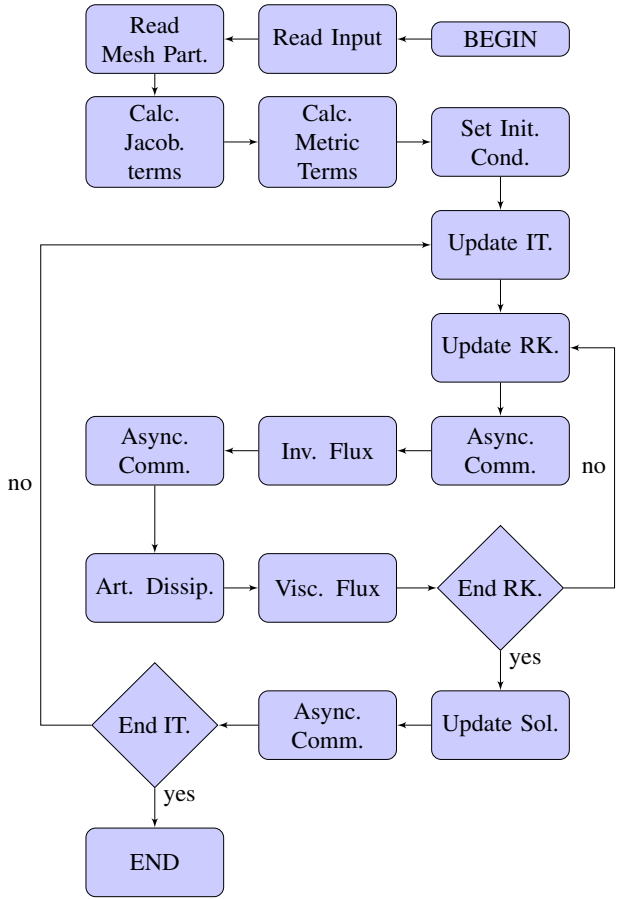

JAZzY is the LES solver used in the current work. A brief 
overview of the code is presented as a flow chart 
illustrated in Fig.\ \ref{fig:jazzy}. Initially, every 
MPI partition reads the same ASCII file which provides 
input data such as flow configurations and simulation 
settings. In the sequence, each MPI partition reads its 
corresponding CGNS mesh file. The Jacobian and the metric 
terms are calculated after the I/O procedure. Then, each 
MPI rank sets the initial conditions defined in the input 
data. The Runge-Kutta time integration is the first routine 
to be called into the interaction loop. 

At the beginning of the time marching scheme, for each sub-step, 
asynchronous communications of the conservative property vector 
are performed, in both, azimuthal and axial directions, followed by 
an update of boundary conditions and dynamic viscosity coefficient. 
Viscous terms are communicated before the computation of artificial 
dissipation operators and viscous flux vectors. MPI waiting functions
are carefully added along the code to avoid out-dated 
information and to enforce the preservation of the numerical method 
accuracy.
Finally, when the requested number of time steps is achieved, 
each MPI partition appends the solution to the output CGNS file. 



\subsection{Inter-Partition Data Exchange}

\begin{figure}[htb]
  \begin{center}
    {\includegraphics[width=0.45\textwidth,trim=0.25cm 0.25cm 0.25cm 0.25cm,clip]
	{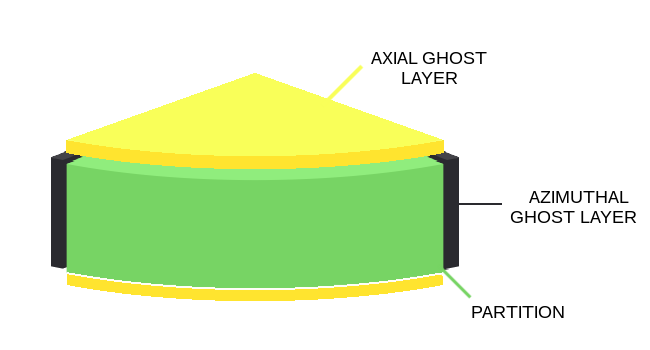}}\\
    \caption{Ghost points creation procedure. The green volume indicates one given
	partition of the domain, while the yellow and black layers represent the axial
    and azimuthal ghost points layers, respectively.}\label{fig:ghost}
  \end{center}
\end{figure}

In the present work, data exchanges are performed using ghost points
which are added to the boundaries of local mesh partitions at the 
main flow direction and at the azimuthal direction in order to carry 
information of the neighboring points. The artificial dissipation 
scheme implemented in the code \citep{jameson_mavriplis_86} uses a 
five-point stencil which demands information of two neighbors, in 
each side, of a given mesh point. Hence, a two-layer ghost point 
fringe is created at the beginning and at the end of each partition. 
Figure \ref{fig:ghost} illustrates the ghost points of a partition 
domain. The yellow and black layers represent the axial and azimuthal 
ghost points, respectively, while the green region is the partition 
mesh.

\begin{figure}[htb]
  \begin{center}
	\subfigure[Forward blocking communication.]{
    \includegraphics[width=0.475\textwidth]{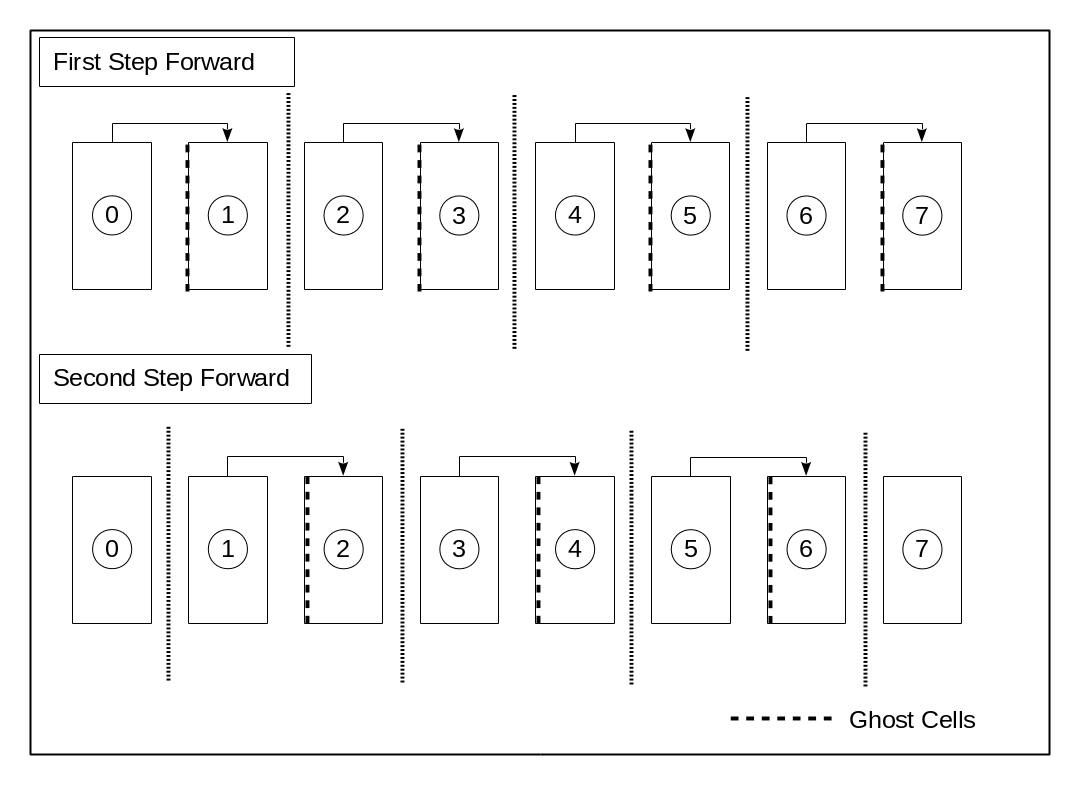}
    \label{fig:forward}
    }
	\subfigure[Backward blocking communication.]{
    \includegraphics[width=0.475\textwidth]{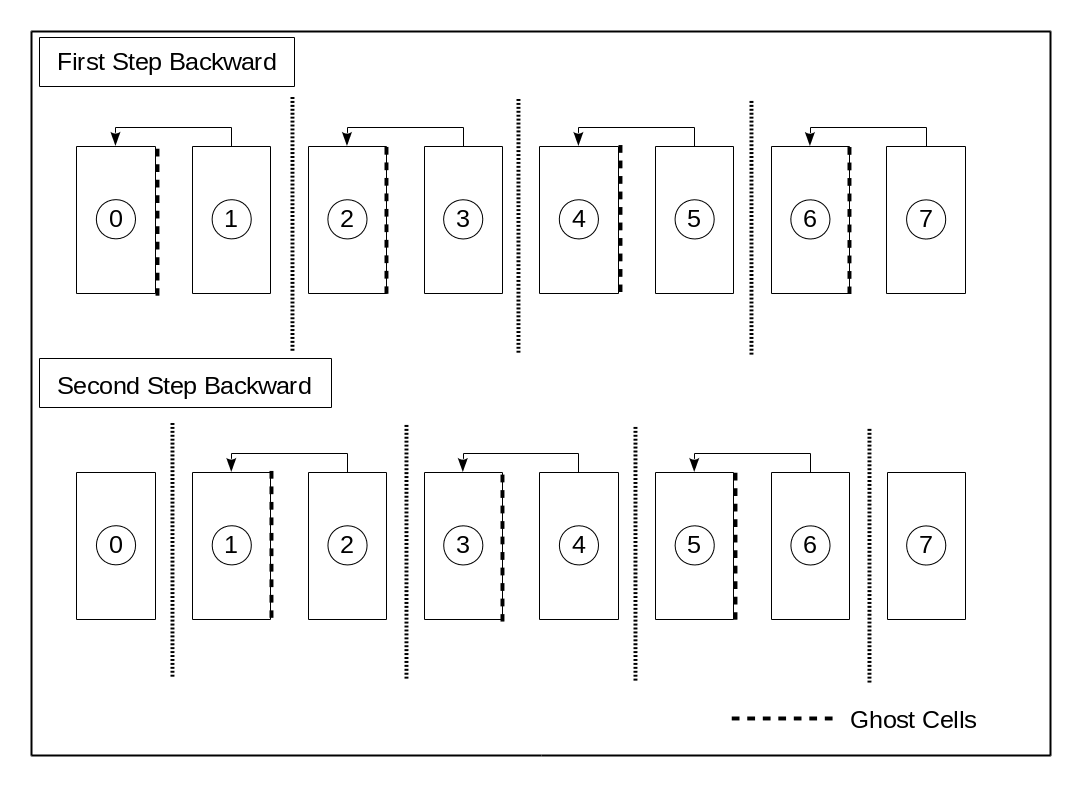}
    \label{fig:backward}
    }
	\caption{Forward and backward communication 
	approaches originally implemented in the code in order 
	to exchange information between neighboring partitions.
    }\label{fig:comm-hpc}
  \end{center}
\end{figure}

The solver was originally developed using a four-step
blocking communication for both axial and azimuthal directions.
Figures \ref{fig:forward} and \ref{fig:backward} demonstrate the 
data exchange approach for a given direction. Initially, the 
communication is performed in the forward direction where even 
partitions send information of their two last local layers to 
the ghost points at the left of odd partitions. If the last 
partition is even, it does not share information in this step. 
In the sequence, odd partitions send information of their two 
last local layers to the ghost points at the left of even 
partitions. If the last partition is odd, it does not share 
information in this step. The third and fourth steps are 
backward communications which are adjusted to work in the same 
fashion as the forward communication does.

The four-step blocking communications are replaced by asynchronous
MPI protocols in order to improve the performance of the solver. 
Waiting functions are carefully added along the Runge-Kutta 
calculations in order to avoid the presence of outdated information 
into the ghost cells and to preserve the accuracy of the time 
integration method.

The meshes used in the current research have a singularity at the 
centerline. It is necessary to correctly treat this region for the 
sake of data consistency. Therefore, properties are extrapolated to 
the singularity in the radial direction and, in the sequence, the master 
partition collects all data from the partitions that share the same 
singularity point and allocates such information into one single vector. 
After the allocation, the properties are averaged in a sequential fashion 
and the result is scattered to the neighbors in the azimuthal direction. 
Such procedure does not use collective communications and it is 
performed in a blocking communication fashion in order to preserve 
the commutative property during the averaging. 
This blocking 
communication is very important in order to achieve the binary 
reproducibility of the computational tool \citep{balaji2013}. This 
approach assures that parallel computations treat the centerline 
singularity in the exactly same fashion sequential calculations do. 
The use of such communication is motivated by the work of Arteage 
{\em et. al.} \cite{arteaga14}. Moreover, such approach can help 
developers verify whether new parallel features implemented into 
the code present the same behavior when running sequentially.
The singularity treatment is the only blocking data 
exchange present into the LES solver after the last upgrades.




\section{Computational Performance Study}\label{sec:comp-perfo}

Large eddy simulations require a significant amount of computational 
resources in order to produce high-quality results. Therefore, it is
very important to evaluate the parallel tool to be sure that it is 
capable of using parallel resources efficiently. Data exchange
between partitions is not free and it can affect the parallel 
computational performance. 

\subsection{Scalability Setup}

Isothermal perfectly expanded jet flow simulations 
are performed using different grid sizes and different partition 
configurations. The Reynolds number of the jet is 
$1.5744 \times 10^{6}$ for the present simulations and a flat-hat 
profile with Mach number of 1.4 is imposed at the
jet exit of the nozzle. Isothermal and zero-velocity 
conditions, along with a zero-pressure-gradient state, consistent
with the freestream condition, are used as initial conditions for 
the simulations. The JAZzY solver has already been validated and 
presented good results using such flow configuration 
\citep{Junior18-abcm}. The reader can find more details about this 
flow configuration in the {\it Compressible Jet Flow Simulation} 
section. In the present work, each simulation performs 1000 
iterations or 24 hours of computation, whatever is shorter in 
terms of computational time. An average of the CPU time per 
iteration through the simulations is measured in order to evaluate 
the weak and strong scalability of the solver using the Santos 
Dumont supercomputer.

One geometry is created for the computational evaluation, where 
the 2-D surface of this computational domain, as presented in 
Fig.\ \ref{fig:bc-1}, is 30 dimensionless units in length and 10 
dimensionless units in height. The diameter of the
jet exit is the length reference unit. In the present work, 13 grids of 
different sizes are created, starting with the coarsest mesh with 
370,000 points up to the finest mesh, with approximately 1 billion 
grid points, as indicated in Tab.\ \ref{tab:mesh}. The mesh size 
growth of the first 12 grids follows a geometric progression 
with ratio 2. 
\begin{table}[htb]
\small\sf\centering
\caption{Mesh configurations used for the scalability study.}\label{tab:mesh}
\begin{tabular}{rrrrc}
\toprule
Mesh & Nb.\ Pt.\ $\xi$ & Nb.\ Pt.\ $\eta$  & 
Nb.\ Pt.\ $\zeta$ & Nb.\ Pt. \\
\midrule
1  & 32   & 32   & 361  & $\approx 370k$\\ 
2  & 64   & 32   & 361  & $\approx 740k$\\ 
3  & 64   & 64   & 361  & $\approx 1.5M$\\ 
4  & 128  & 64   & 361  & $\approx 3.0M$\\ 
5  & 128  & 128  & 361  & $\approx 6.0M$\\ 
6  & 256  & 128  & 361  & $\approx 11.8M$\\ 
7  & 256  & 256  & 361  & $\approx 23.7M$\\ 
8  & 512  & 256  & 361  & $\approx 47.3M$\\ 
9  & 512  & 512  & 361  & $\approx 94.6M$\\ 
10 & 1024 & 512  & 361  & $\approx 190M$\\ 
11 & 1024 & 1024 & 361  & $\approx 380M$\\ 
12 & 2048 & 1024 & 361  & $\approx 760M$\\ 
13 & 1700 & 1700 & 361  & $\approx 1.0B$\\ 
\bottomrule
\end{tabular}
\end{table}

Simulations performed in the present study are run using up to 3072 
computational cores. Different partitioning configurations are 
evaluated for a given number of processors. Table \ref{tab:part} 
presents the number of partitions in the azimuthal direction, NPZ, 
for a given number of computational cores.
\begin{table}[htb]
\small\sf\centering
\caption{Grid partitioning configurations for different number of 
cores.}\label{tab:part}
\begin{tabular}{r|rrrrrr}
\toprule
Nb. Cores & {\bf NPZ} \\ 
\midrule
2 & 1 & 2 \\ 
4 & 1 & 2 & 4 \\ 
8 & 1 & 2 & 4 & 8 \\ 
16 & 1 & 2 & 4 & 8 & 16 \\ 
32 & 1 & 2 & 4 & 8 & 16 & 32 \\ 
64 & 2 & 4 & 8 & 16 & 32 \\ 
128 & 2 & 4 & 8 & 16 & 32 \\ 
256 & 4 & 8 & 16 & 32 \\ 
512 & 4 & 8 & 16 & 32 \\ 
1024 & 8 & 16 & 32 \\ 
2048 & 8 & 16 & 32 \\ 
3072 & 24 & 48 & 96 \\ 
\bottomrule
\end{tabular}
\end{table}
The partitioning configurations which provided the fastest 
calculation are used to evaluate the scalability of the 
solver.
Table \ref{tab:ratio} presents the number of ghost points
as a percentage of the number of grid points, from the optimal
partitioning configuration, for all mesh configurations and 
computational cores used in the strong scaling study.
One can understand the ghost point per total grid point 
ratio as the communication cost for a parallel calculation 
using MPI. In the present work, such cost is, for most of 
the partitioning configurations, less than 1\% and
approximately 7\% in the worst case scenario.
\begin{table*}[thb!]
\small\sf\centering
\caption{Number of ghost points as a percentage of 
the total number of grid points for every grid and amount of
computational resources used in the strong scaling 
study.}\label{tab:ratio}
\begin{tabular}{r|rrrrrrrrrrrr}
\toprule
{\tiny Mesh} & {\tiny Nb. Cores} \\
	 & 2     & 4      & 8      & 16     & 32     & 64     & 128    & 256    & 512    & 1024   & 2048   & 3072 \\  
\midrule
1	& \scriptsize 0.81\% & \scriptsize 0.52\% & \scriptsize 0.66\% & \scriptsize 0.93\%	& \scriptsize 1.46\% & \scriptsize 1.82\% & \scriptsize 2.53\% & \scriptsize 3.96\% & \scriptsize 6.80\% \\
2	& \scriptsize 0.26\% & \scriptsize 0.33\% & \scriptsize 0.47\% & \scriptsize 0.74\%	& \scriptsize 0.93\% & \scriptsize 1.46\% & \scriptsize 1.82\% & \scriptsize 2.53\% & \scriptsize 3.96\% \\
3	& \scriptsize 0.21\% & \scriptsize 0.17\% & \scriptsize 0.23\% & \scriptsize 0.37\%	& \scriptsize 0.64\% & \scriptsize 0.73\% & \scriptsize 0.91\% & \scriptsize 1.27\% & \scriptsize 1.98\% \\
4	& \scriptsize 0.08\% & \scriptsize 0.13\% & \scriptsize 0.17\% & \scriptsize 0.23\%	& \scriptsize 0.60\% & \scriptsize 0.64\% & \scriptsize 0.73\% & \scriptsize 0.91\% & \scriptsize 1.27\% \\
5	& \scriptsize 0.04\% & \scriptsize 0.07\% & \scriptsize 0.08\% & \scriptsize 0.16\%	& \scriptsize 0.16\% & \scriptsize 0.32\% & \scriptsize 0.33\% & \scriptsize 1.58\% & \scriptsize 1.60\% & \scriptsize 1.63\% \\
6	& \scriptsize 0.03\% & \scriptsize 0.04\% & \scriptsize 0.06\% & \scriptsize 0.08\%	& \scriptsize 0.13\% & \scriptsize 0.30\% & \scriptsize 0.23\% & \scriptsize 0.45\% & \scriptsize 1.58\% &  \scriptsize 1.60\% \\
7	& \scriptsize 0.02\% & \scriptsize 0.02\% & \scriptsize 0.04\% & \scriptsize 0.08\%	& \scriptsize 0.06\% & \scriptsize 0.08\% & \scriptsize 0.12\% & \scriptsize 0.18\% & \scriptsize 0.23\% &  \scriptsize 0.80\% \\
8	& \scriptsize 0.01\% & \scriptsize 0.02\% & \scriptsize 0.03\% & \scriptsize 0.04\%	& \scriptsize 0.05\% & \scriptsize 0.06\% & \scriptsize 0.08\% & \scriptsize 0.13\% & \scriptsize 0.23\% &  \scriptsize 0.42\% \\
9	& \scriptsize        & \scriptsize        & \scriptsize        & \scriptsize        & \scriptsize 0.02\% & \scriptsize 0.03\% & \scriptsize 0.05\% & \scriptsize 0.11\% & \scriptsize 0.09\% &  \scriptsize 0.21\% & \scriptsize 0.40\% & \scriptsize 0.16\% \\
10	& \scriptsize        & \scriptsize        & \scriptsize        & \scriptsize        & \scriptsize        & \scriptsize 0.02\% & \scriptsize 0.04\% & \scriptsize 0.05\% & \scriptsize 0.04\% &  \scriptsize 0.09\% & \scriptsize 0.21\% & \scriptsize 0.13\% \\
11	& \scriptsize        & \scriptsize        & \scriptsize        & \scriptsize        & \scriptsize        & \scriptsize        & \scriptsize 0.02\% & \scriptsize 0.02\% & \scriptsize 0.04\% &  \scriptsize 0.05\% & \scriptsize 0.06\% & \scriptsize 0.07\% \\
12	& \scriptsize        & \scriptsize        & \scriptsize        & \scriptsize        & \scriptsize        & \scriptsize        &  \scriptsize       & \scriptsize 0.02\% & \scriptsize 0.01\% &  \scriptsize 0.04\% & \scriptsize 0.05\% & \scriptsize 0.06\% \\
13	& \scriptsize        & \scriptsize        & \scriptsize        & \scriptsize        & \scriptsize        & \scriptsize        &  \scriptsize       & \scriptsize 0.01\% & \scriptsize 0.02\% &  \scriptsize 0.02\% & \scriptsize 0.03\% & \scriptsize 0.03\% \\
\bottomrule
\end{tabular}
\end{table*}


\subsection{Strong Scaling}

The speedup is one of the most common figures of  
metric for the performance evaluation of parallel algorithms 
and architectures \citep{Ertel94} and it is used in the present 
work in order to measure the strong scaling of the solver and 
compare it with the ideal case. Different approaches are used 
by the scientific community in order to calculate the speedup 
\citep{xian10,gustafson88}. In the present work the speedup, 
${Sp}({\bf m},N)$, is given by 
\begin{equation}
	{S_p}({\bf m},N) = \frac{T({\bf m},s)}{T({\bf m},N)} \, \mbox{.}
	\label{eq:speedup}
\end{equation}
In this expression, $T$ stands for the time spent by ${\bf m}$-th mesh 
to perform one thousand time steps, $N$ represents the number of 
computational cores and $s$ is the starting point of the scalability 
study. The strong scaling efficiency of a given mesh configuration, 
as a function of the number of processors, is written considering 
the law of Amdahl \cite{amdahl67} as
\begin{equation}
	\eta^{st}({\bf m},N) = \frac{{Sp}({\bf m},N)}{N} \, \mbox{.}
	\label{eq:eff}
\end{equation}
Ideally, the sequential computation is the starting point for a 
strong scalability study. However, quite often, the computational 
problem is too big to fit in one single cluster node due to hardware 
limitations of the computer node. Hence, it is necessary to shift 
the starting point to a minimum number of processors in which the 
code can be run. 
For such cases, where $N=s$, speedup 
and strong scaling efficiency are assumed as
\begin{eqnarray}
	{S_p}({\bf m},s) = s & \mbox{and} & 
	\eta^{st}({\bf m},s) = 1\, \mbox{.}
	\label{eq:starting-pt}
\end{eqnarray}
Table \ref{tab:strong} presents the minimum and maximum number of 
computational cores used for each mesh configuration in the strong 
scaling study. 
\begin{table}[htb]
\small\sf\centering
\caption{Starting point, $s$, and maximum number of computational 
cores, $N_{max}$, used by each grid configuration.}\label{tab:strong}
\begin{tabular}{rrr|rrr}
\toprule
Mesh & $s$ & $N_{max}$ & Mesh & $s$ & $N_{max}$\\
\midrule
1  & 1   & 512  & 8  & 1   & 1024 \\
2  & 1   & 512  & 9  & 32  & 3072 \\
3  & 1   & 512  & 10 & 64  & 3072 \\
4  & 1   & 512  & 11 & 128 & 3072 \\
5  & 1   & 1024 & 12 & 256 & 3072 \\
6  & 1   & 1024 & 13 & 256 & 3072 \\
7  & 1   & 1024 \\ 
\bottomrule
\end{tabular}
\end{table}


\subsubsection{Evaluation of Code Improvements}

An initial evaluation of the improvements brought to the code
is performed in a smaller supercomputer before studying
the scalability of the new version of the solver in the
Santos Dumont supercomputer. Optimal workload distribution
and non-blocking communications are the main modifications
delivered to the LES solver. A strong scalability comparison 
is fulfilled using the Euler supercomputer which is included 
into a national project known as CEPID-CeMEAI 
\cite{cepid-euler}. The cluster is an Hewlett Packard 
Enterprise (HPE) machine. It presents 104 computational nodes 
and each one has two deca-core 2.8 GHz 
{\it Intel Xeon\textsuperscript{\textregistered}} \, 
E52680v2 processors and 128 Gb DDR3 1866MHz random access 
memory. The entire cluster has 2080 computational cores 
available for the project members. 

\begin{figure}[htb!]
  \begin{center}
	\includegraphics[width=0.45\textwidth,trim=0.5cm 0.5cm 0.5cm 0.5cm,clip]{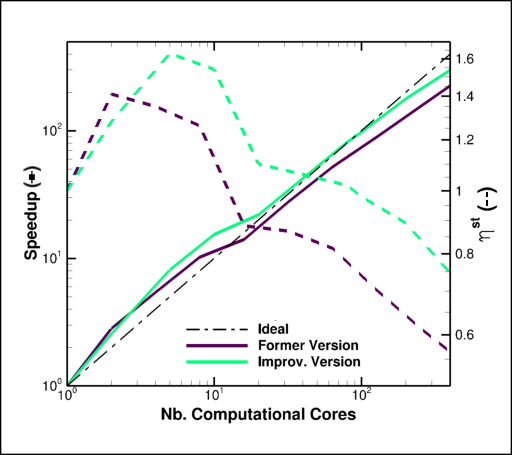} 
	\caption{Comparison of speedup (\textbf{--}) and 
    strong scaling efficiency (\textbf{- -}) curves from both 
    versions of the JAZzY solver over  approximately 50 million 
    grid points, {\it i.e.}, mesh 8.}
    \label{fig:improv}
  \end{center}
\end{figure}

Simulations are run using mesh 8 configuration, which has
$\approx 50$ million grid points, as indicated in Tab.\ 
\ref{tab:mesh}. The study is performed allocating up to 400 
computational cores in parallel. Figure \ref{fig:improv} presents 
a comparison between the strong scalability from both versions 
of the code. One can observe that the strong efficiency curve is 
shifted up approximately by 20\% with the improvements implemented 
in the code for the grid configuration evaluated. The results 
enhance the importance of the latest upgrades brought to the 
JAZzY solver.



\subsubsection{Strong Scalability on Santos 
Dumont Supercomputer}

After the evaluation of recent modifications,
a more detailed scalability study is done using the Teraflop
branch from the Santos Dumont supercomputer.
The LES solver, used in the present study, can run up to 
$\approx 50$ Million grid points, {\em i.e.}, mesh 8 configuration, 
in one single node of the Santos Dumont machine, which has 
64GB of RAM. The standard maximum number of computational 
cores allowed to be used for scalability tests in the Brazilian 
computer is 3072. More than 400 simulations are run in order to 
perform the scalability study, considering all mesh configurations 
and different partitioning arrangements. 

\begin{figure}[htb!]
  \begin{center}
	\includegraphics[width=0.45\textwidth,trim=0.5cm 0.5cm 0.5cm 0.5cm,clip]{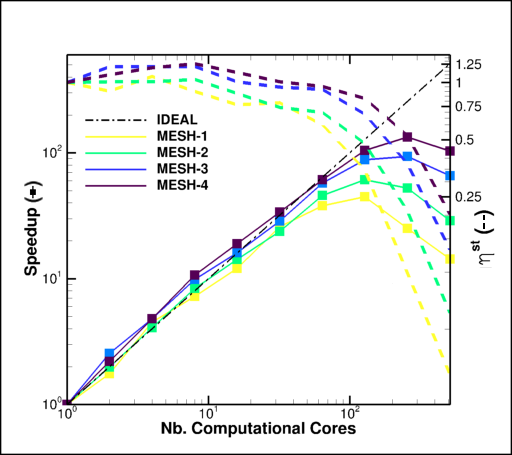} 
	\caption{Speedup (\textbf{--}) and strong efficiency (\textbf{- -})
    of meshes 1, 2, 3 and 4.}
    \label{fig:st-1-4}
  \end{center}
\end{figure}

\begin{figure}[htb!]
  \begin{center}
	\includegraphics[width=0.45\textwidth,trim=0.5cm 0.5cm 0.5cm 0.5cm,clip]{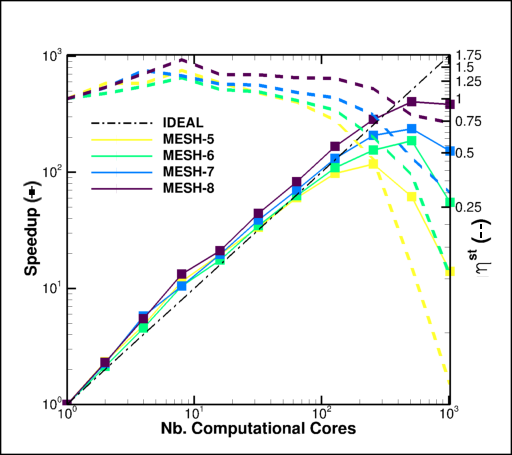} 
	\caption{Speedup (\textbf{--}) and strong efficiency (\textbf{- -})
    of meshes 5, 6, 7 and 8.}
    \label{fig:st-5-8}
  \end{center}
\end{figure}

\begin{figure}[htb!]
  \begin{center}
	\includegraphics[width=0.45\textwidth,trim=0.5cm 0.5cm 0.5cm 0.5cm,clip]{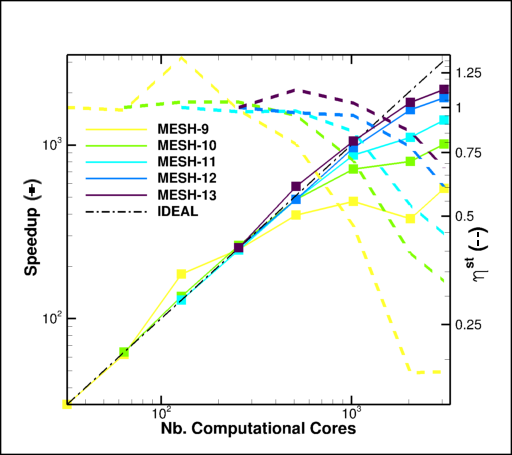} 
	\caption{Speedup (\textbf{--}) and strong efficiency (\textbf{- -})
    of meshes 9, 10, 11, 12 and 13.}
    \label{fig:st-9-13}
  \end{center}
\end{figure}

Figures \ref{fig:st-1-4}, \ref{fig:st-5-8} and \ref{fig:st-9-13} 
present the speedup, in solid line, and the strong efficiency,
in dashed line, for the 13 mesh configurations.
One can notice that the speedup and strong efficiency of all 
numerical test cases present similar trends. The speedup 
proportionally increases with the number of processors until it 
reaches a saturation point, which indicates that MPI data 
exchange is becoming as time consuming as the calculations 
themselves. This fact is reinforced by the value of
the strong scalability efficiency of approximately 50\% at 
the vicinity of the speedup peak, for all meshes evaluated. 

The two smallest problems evaluated in the present paper present 
similar strong scaling behavior. In both cases, the scalability 
curves follow the ideal case for $N \leq 8$ and they reach a maximum 
speedup of approximately 60 using 128 computational cores. The 
results with Meshes 3 and 4 indicate a higher value of speedup peak, 
which is approximately 100 for $N=256$, when compared to the two 
smallest problems. Moreover, they present super-scalability when 
running with less than 16 partitions in parallel and 
$\eta^{st}\approx 1$ for $32<N<64$. 

A similar shape for the speedup curves, as observed in Fig.\ 
\ref{fig:st-1-4}, can also be seen in Fig.\ \ref{fig:st-5-8}. 
However, the speedup saturation point is shifted upwards with the 
increase in the size of the computational problem. 
Mesh 5 configuration presents a maximum speedup of $\approx 100$ 
for $N=256$, while Mesh 6 test case has a speedup peak of 
$\approx 200$ when running on 512 cores in parallel. Both cases 
have super-linear scaling for $N \leq 32$ and strong efficiency of 
$100\%$ when running on 64 MPI partitions. Meshes 7 
and 8 present the speedup apex of $\approx 250$ and 
$\approx 500$, respectively, for $N=512$. The former case 
indicates $\eta^{st}\geq 1$ for $N<256$, while the latter 
presents super-scaling for $N\leq 512$.

The last five cases, presented in Fig.\ \ref{fig:st-9-13}, which 
have more than 90 million grid points, indicate better strong 
scaling when compared to the previous test cases. For these 
five cases, the speedup scales proportionally to the number of 
available resources. Meshes 9, 10 and 11 present 
$\eta^{st} \geq 1$ for $N \leq 256$, although the efficiency 
curves are reasonably flat and around $\eta^{st} \approx 1$ for 
meshes 10 and 11 all the way up to $N \leq 1024$.
The largest grid size analyzed here, {\em i.e.}, mesh 13, 
presents $\eta^{st} \approx 1$ for calculations performed using 
up to 2048 partitions and high strong efficiency, 
$\eta^{st} \approx 70\% $, when using 3072 computational cores 
in parallel. 

The presence of super-linear scalability, as observed in Figs.\ 
\ref{fig:st-1-4}, \ref{fig:st-5-8} and \ref{fig:st-9-13}, can be 
explained by fact that cache memory no longer becomes a bottleneck 
with the increase of the number of computational resources. Since 
more cache memory is available, when the number of processors is 
increased, it can be used more effectively by the workload in 
each process when compared to the use of cache memory of
starting point calculation of the strong scalability study 
\citep{Ristov16,Gusev14}. 
The cache memory bottleneck could be related to a non-optimal 
memory access design of the code. Furthermore, it could be 
correlated to the deterioration of performance when increasing 
the number of processors used in a calculation. The code 
would presumably not deliver optimal efficiency when using 
hundreds of thousands of cores in parallel. The developers 
are nevertheless investigating the cause of such issues. 
Implementing cache blocking \cite{Jin16,Kamil05,Gannon87}
and vectorization \cite{Lohner10,Basermann-et-al-10}
techniques could overcome the difficulty simultaneously 
with an eventual further improvement of the solver performance 
in parallel.

It is important to remark that the authors are interested on 
simulations of supersonic jet flows. Calculations of this flow 
configuration, using the LES formulation, require meshes with 
more than 100 million points. The present study indicates that 
the JAZzY solver has a good performance for problems of this 
size and it is capable of using the Santos Dumont supercomputer 
efficiently. Moreover, the strong scaling evaluation can be used 
as a guideline for the selection of the best partitioning 
configuration and/or the number of processors to be used in the 
Brazilian cluster, for a given mesh size required for the 
particular study.



\subsection{Weak Scalability on Santos 
Dumont Supercomputer}

Weak scalability can be interpreted as the ability of conserving the computing
time for a fixed workload. In the ideal case, a parallel code should preserve 
constant time-to-solution when the dimension of a problem increases at the 
same rate as the number of computational resources. The weak scaling 
efficiency for a given workload, $w$, can be written mathematically as
\begin{equation}
	\eta^{wk}(w,N) = \frac{T(w,s)}{T(w,N)} \, \mbox{.}
	\label{eq:weak-eff}
\end{equation}

The weak scaling of the parallel solver is evaluated using 5 different 
workloads, which are represented in the present paper by the number of grid 
points divided by the number computational cores. Table \ref{tab:weak} 
presents the mesh and the number of processing units used at the starting 
point followed by the maximum number of computing units and the workload for 
each weak scalability study. 
\begin{table}[htb]
\small\sf\centering
\caption{Description of workloads and starting points used for the
weak scalability study.}\label{tab:weak}
\begin{tabular}{lllll}
\toprule
Case & 1st Mesh & $s$ & $N_{max}$ &  Workload 
{\scriptsize $\left[ \frac{\mbox{ Nb.\ Pt.\ }}{\mbox{Core}} \right]$} \\
\midrule
A  & Mesh 1 & 2 & 2048 & $\approx185k$ \\ 
B  & Mesh 1 & 1 & 2048 & $\approx370k$ \\ 
C  & Mesh 2 & 1 & 2048 & $\approx740k$ \\ 
D  & Mesh 3 & 1 & 1024 & $\approx1.5M$ \\ 
E  & Mesh 4 & 1 &  512 & $\approx3.0M$ \\ 
\bottomrule
\end{tabular}
\end{table}
The computational grids, presented in Tab.\ \ref{tab:mesh}, are created using a 
total number of mesh points which is proportional to the power of two in order to 
be able to keep a constant workload when the computational resources of a given test 
case is doubled. However, due to limitations of the parallel tool, Mesh 13 configuration
is an exception to such trend and presents $\approx 1.0B$ points, which is $\approx 1.45$ 
the size of Mesh 12 test case that has $\approx 760M$ points. Therefore, for this special 
grid configuration, a correction is used on the calculation of $\eta^{wk}$ in 
order to properly compare it with data collected from other mesh configurations. It 
is also important to remark here that, the first case is the only one to use two 
computational cores as a starting point, $s$. The other weak scaling studies are 
performed using a sequential computation as a starting point.

\begin{figure}[htb!]
  \begin{center}
	\includegraphics[width=0.45\textwidth,trim=0.5cm 0.5cm 0.5cm 0.5cm,clip]{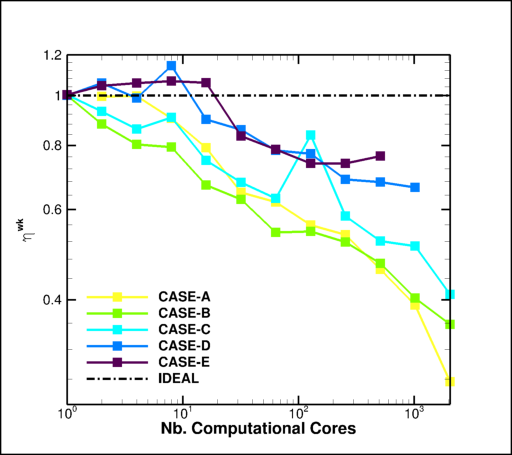} 
    \caption{Weak scalability of the JAZzY solver running on the Santos Dumont 
	supercomputer.}\label{fig:weak}
  \end{center}
\end{figure}

Figure \ref{fig:weak} presents the weak scalability of the five different 
workloads evaluated in the current paper. 
In the worst case scenario, the parallel code can still present a good weak scalability 
with $\eta^{wk}\approx 0.7$ for the two largest workloads evaluated in present work, 
1.5M and 3.0M grid points per core. Nonetheless, one can notice the weak scaling 
efficiency decay when increasing the number of processors used on the calculations. 
Moreover, the super-linear speedup behavior, previously observed on the strong scaling 
study, matches with weak scalability peaks. One example is the sudden rise of 
$\eta^{wk}$ present on the Case {\it C} scaling illustrated in Fig.\ \ref{fig:weak} 
for 128 computational cores. Such event matches the strong super-scalability 
behavior of Mesh 9 test case for 128 processing units indicated in Fig.\ \ref{fig:st-9-13}.




\section{Compressible Jet Flow Simulation}\label{sec:val}

This section presents results achieved from the simulation of a 
supersonic jet flow configuration. This calculation is performed 
in order to validate the LES code, and it is included here simply 
to demonstrate that the numerical tool is indeed capable of 
presenting physically sound results for the problem of interest. 
Results are compared to numerical \citep{Mendez12} and to 
experimental data \citep{bridges2008turbulence}. The reader can 
find more details of this particular simulation in the work of 
{\em Junqueira-Junior et. al.} \cite{Junior18-abcm}. 

\begin{figure}[htb!]
  \begin{center}
    \subfigure[2-D cut of the geometry colored by velocity magnitude 
	contours.]{
    \includegraphics[trim= 5mm 5mm 5mm 5mm,clip,width=0.4\textwidth]
	{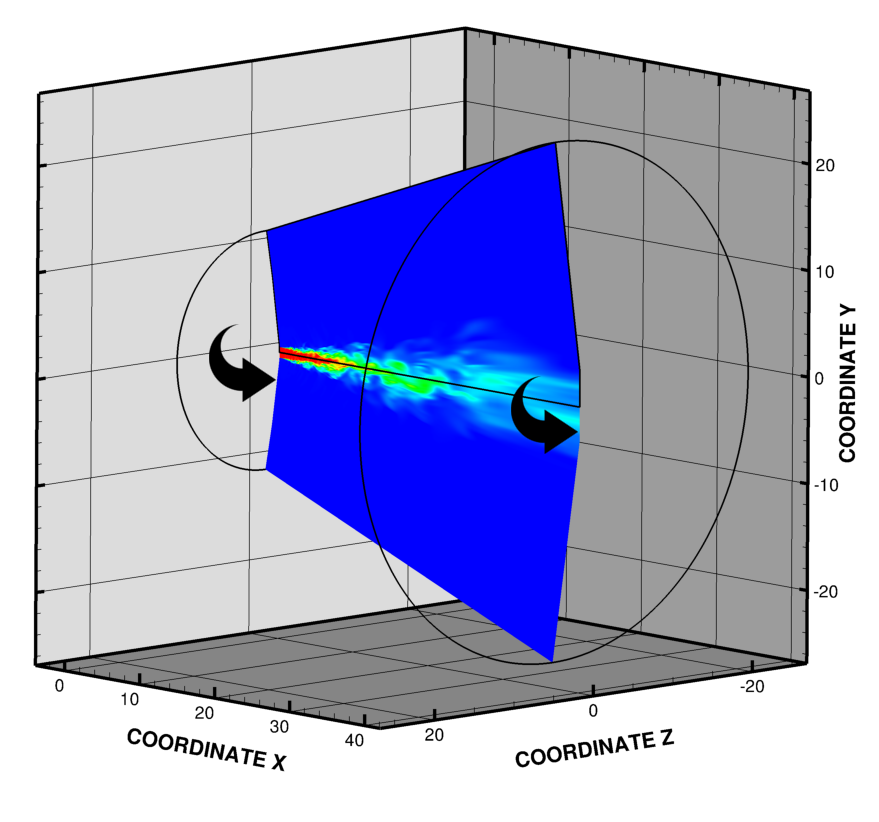} 
    \label{fig:mesh}
    }
	\subfigure[2-D cut of the domain superimposed by grid points 
	distribution.]{
    \includegraphics[trim= 5mm 5mm 5mm 5mm,clip,width=0.45\textwidth]
	{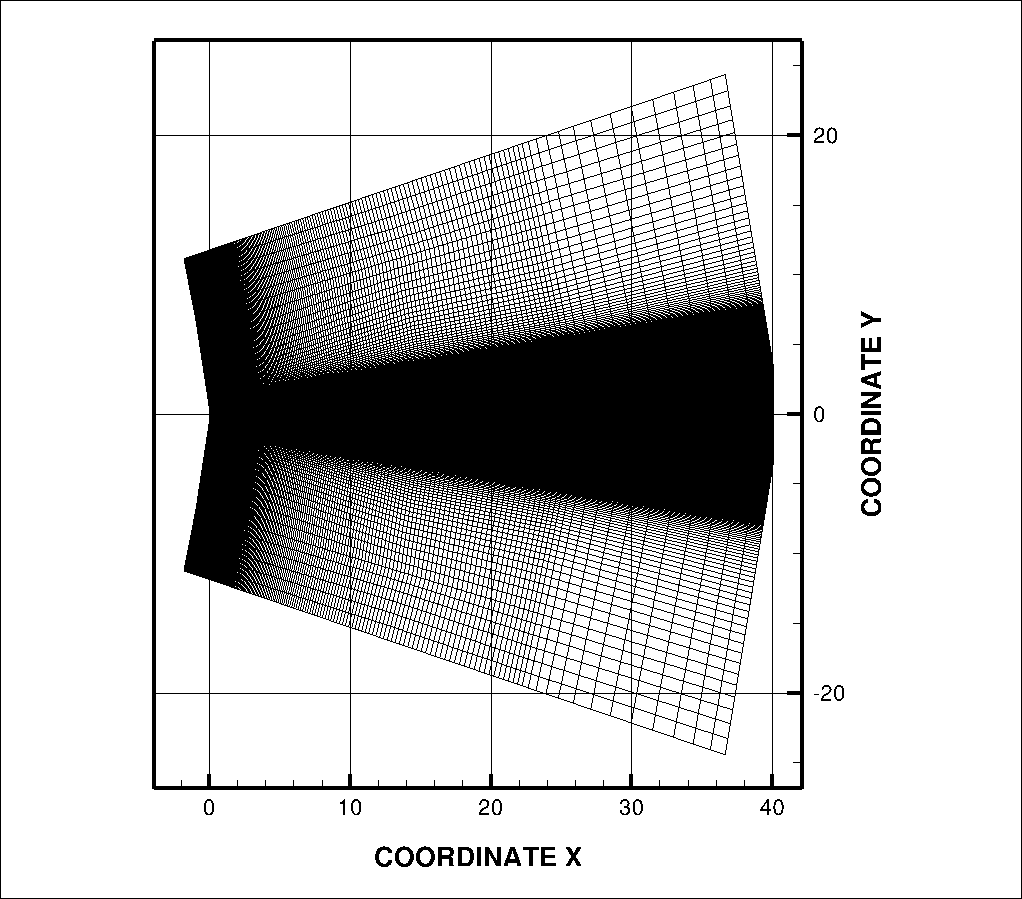}
    \label{fig:geom-val}
    }
    \caption{Illustration of geometry and mesh used into the 
    validation of the LES solver.}\label{fig:mesh-geom}
  \end{center}
\end{figure}

A geometry is created using a divergent shape whose axis length is 
40 times the diameter of the jet exit, $D$.
The minimum and maximum heights of the domain are $\approx 16D$ 
and $25D$, respectively. Figure \ref{fig:mesh-geom} illustrates 
a 2-D cut of the geometry and the grid point distribution used on 
the validation of the solver. The mesh created to validate the 
parallel solver is composed of 343 points in the axial direction, 
398 points in the radial direction and 360 points in the azimuthal 
direction. This yields a grid with approximately 50 million points. 
The distance between mesh points increases towards the outer region 
of the domain. This procedure forces the dissipation of properties 
far from the jet exit of the nozzle in order to avoid 
reflection of waves into the domain. The grid coarsening can be 
understood as an implicit damping which can smooth out properties 
far from the entrance domain, in the region where 
the mesh is no longer refined. The calculation is performed using 
500 computational cores.

An isothermal perfectly expanded jet flow is studied 
for the present validation. The Mach number at the 
exit of the nozzle is $M = 1.4$\@. The pressure ratio, 
$PR=P_{j}/P_\infty$, and the temperature ratio, 
$TR=T_{j}/T_\infty$, between the entrance of the 
domain and the ambient freestream conditions, are equal to one, 
{\em i.e.}, $PR = 1$ and $TR=1$. The $j$ subscript identifies the 
properties at the jet exit of the nozzle and the 
$\infty$ subscript stands for properties at the farfield region. 
The Reynolds number of the jet is $Re = 1.57 \times 10^{6}$, 
based on the diameter of the domain entrance, $D$. 
The time increment, $\Delta t$, used for the validation study is 
$1\times 10^{-4}$ dimensionless time units. 

The boundary conditions previously presented in the 
{\it Large Eddy Simulation Formulation} section are applied 
in the current simulation. Initial conditions
of the computation are set as isothermal and 
zero-pressure-gradient states, consistent with to the 
freestream condition, along with stagnated velocity, 
{\em i.e.}, zero velocity. 
The simulation runs a predetermined period of time until the 
statistically steady flow condition is achieved. This first 
pre-calculation is important in order to assure that the jet 
is fully developed and turbulent. 

In the present work, the pre-calculation is run for 14 flow 
through time (FTT) units before reaching the statistically 
steady flow condition. One FTT unit represents the necessary 
amount of time for a particle to cross all the domain, in the 
main flow direction, considering the inlet velocity at the jet 
exit of the nozzle. After the statistically stationary state 
is achieved, the simulations are restarted and run for 3.30 
FTT more in which data of the flow are extracted and recorded 
in a frequency of 50Hz\@. 
The data sampling can be as well represented by a Strouhal number 
range of $ 0.01 \lesssim S_{t} \lesssim 36 $. The Strouhal
number, $S_{t}$, is calculated based on the frequency of extraction, 
Mach number at the exit of the nozzle, and the jet diameter.

\begin{figure}[htb!]
  \centering
  \subfigure[Time averaged axial component of velocity, 
  $\langle U \rangle$.]
  {\includegraphics[trim= 5mm 5mm 5mm 5mm, clip, width=0.45\textwidth]{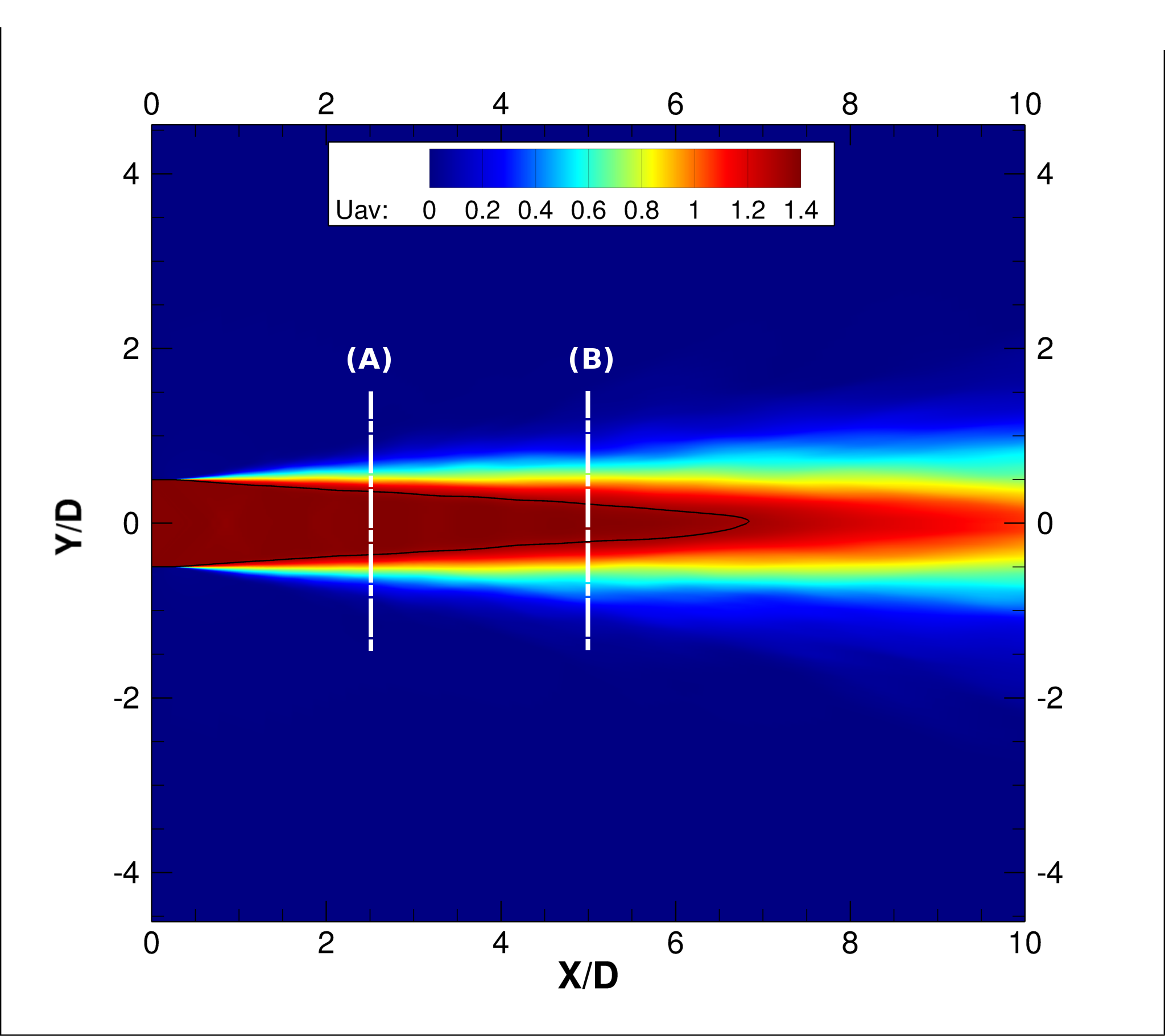}\label{fig:uav}}
  \subfigure[RMS values of the fluctuating part of velocity 
  axial component, $u_{RMS}^{*}$.]
    {\includegraphics[trim= 5mm 5mm 5mm 5mm, clip, width=0.45\textwidth]{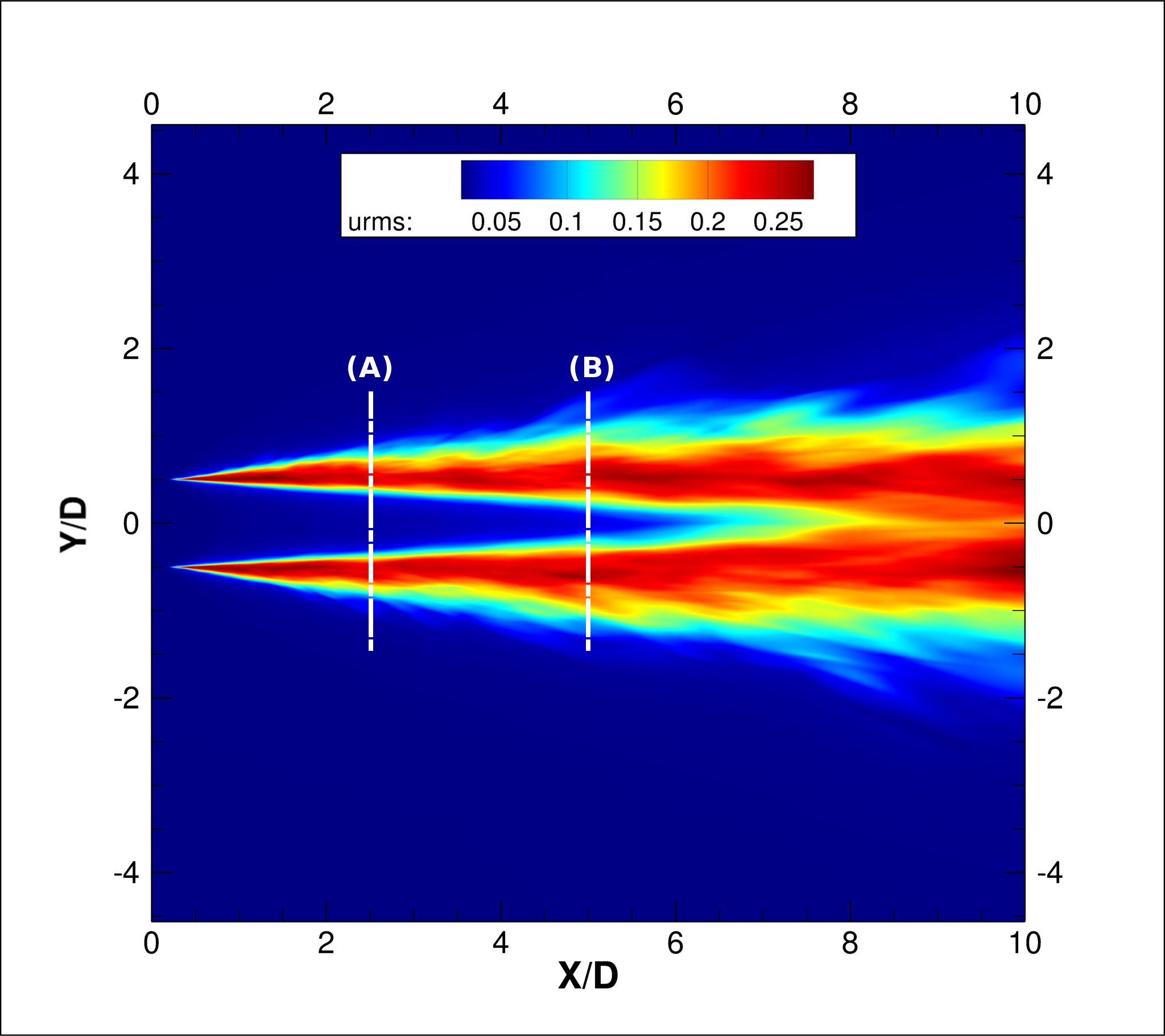}\label{fig:urms}}
	\caption{Lateral view of distributions of 
    $\langle U \rangle$ and $u_{RMS}^{*}$. The white dashed 
    lines indicate the positioning of radial cuts where data 
    are extracted and averaged. The solid black line in 
    {\bf (a)} represents the potential core of the jet.}
	\label{fig:dist-u}
\end{figure}

Figure \ref{fig:dist-u} indicates the positioning of the two surfaces, 
(A) and (B), where data are extracted and averaged through time. Cuts 
(A) and (B) are radial profiles at $2.5D$ and $5.0D$ units downstream 
of the domain entrance. Flow quantities are also averaged in the 
azimuthal direction when the radial profiles are calculated. 
Moreover, 
Figs.\ \ref{fig:uav} and \ref{fig:urms} present distributions 
of time averaged axial component of velocity and root mean square 
values of the fluctuating part of the axial component of velocity, 
which are represented in the present work as $\langle U \rangle$ 
and $u_{RMS}^{*}$, respectively. The solid black line indicated in 
Fig.\ \ref{fig:uav} represents the potential core of the jet, which 
is defined as the region where the time averaged axial velocity 
component is at least $95\%$ of the velocity of the jet at the 
inlet.

\begin{figure}[htb!]
  \centering
    {\includegraphics[trim= 5mm 5mm 5mm 5mm, clip, width=0.45\textwidth]{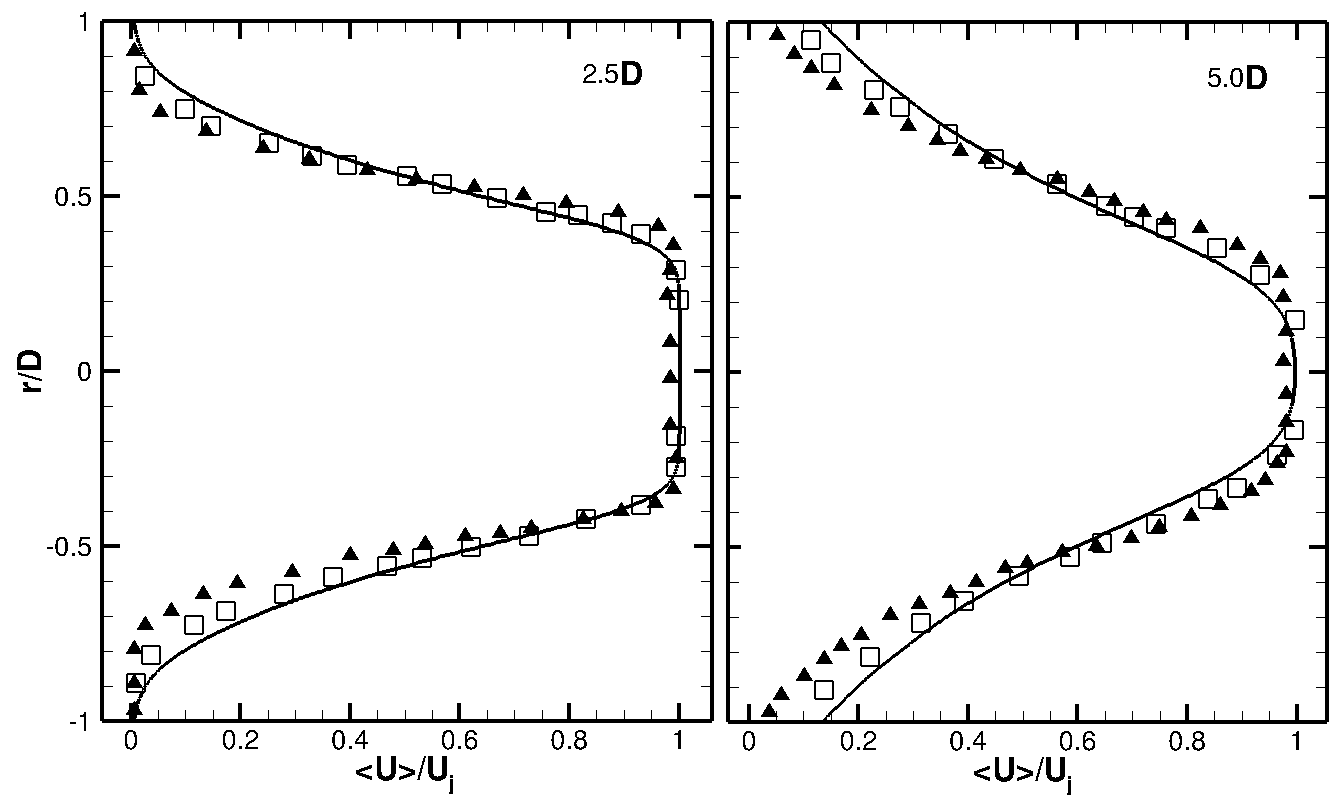}}
    \caption{Profiles of the averaged axial component 
    of velocity, $\langle U \rangle$, at $2.5D$ and $5.0D$ from 
    the entrance: (\textbf{--}) JAZzY results; ($\square$) numerical 
    data; ($\blacktriangle$) experimental data.}
	\label{fig:prof-uav}
\end{figure}

\begin{figure}[htb!]
  \centering
  {\includegraphics[trim= 5mm 5mm 5mm 5mm, clip, width=0.45\textwidth]{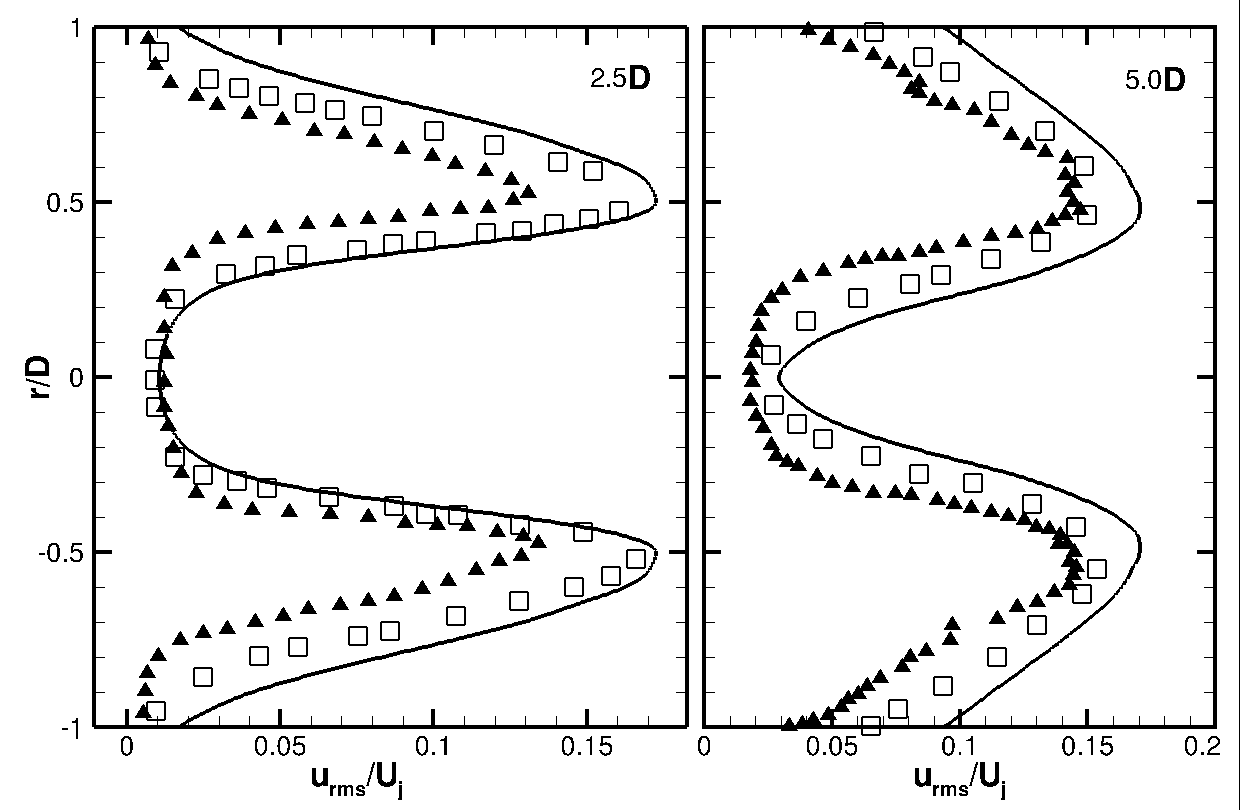}}
  \caption{Profiles of the RMS of the fluctuation part of 
    axial component of velocity, $u_{RMS}^{*}$, at $2.5D$ and $5.0D$ 
    from the entrance: (\textbf{--}) JAZzY results; ($\square$) 
    numerical data; ($\blacktriangle$) experimental data.}
	\label{fig:prof-urms}
\end{figure}

Dimensionless profiles of $\langle U \rangle$ and $u^{*}_{RMS}$ 
at the cuts along the mainstream direction of the computational 
domain are compared with numerical and experimental results in 
Figs.\ \ref{fig:prof-uav} and \ref{fig:prof-urms}, respectively. 
The solid line stands for results achieved using the JAZzY code 
while square and triangular symbols represent numerical 
\citep{Mendez12} and experimental 
\citep{bridges2008turbulence} data, respectively. 
The averaged profiles obtained in the present work correlate 
well with the reference data at the two positions compared here. 
Nevertheless, the results achieved for both the JAZzY solver and 
by the numerical reference \citep{Mendez12} 
present difficulties to correctly predict the peaks of 
$u^{*}_{RMS}/U_{j}$ near $r/D=0.5D$ at $2.5D$ when compared 
with experimental data \citep{bridges2008turbulence}, as 
indicated in Fig.\ \ref{fig:prof-urms}. It is important
to remark that the LES tool can provide good predictions of 
supersonic jet flow configurations when using a sufficiently 
fine grid point distribution. Therefore, efficient massive 
parallel computing is mandatory in order to achieve high-quality
results. 

\begin{figure}[htb!]
  \centering
    {
	  \includegraphics[trim= 1mm 1mm 1mm 1mm, clip, width=0.45\textwidth]
      {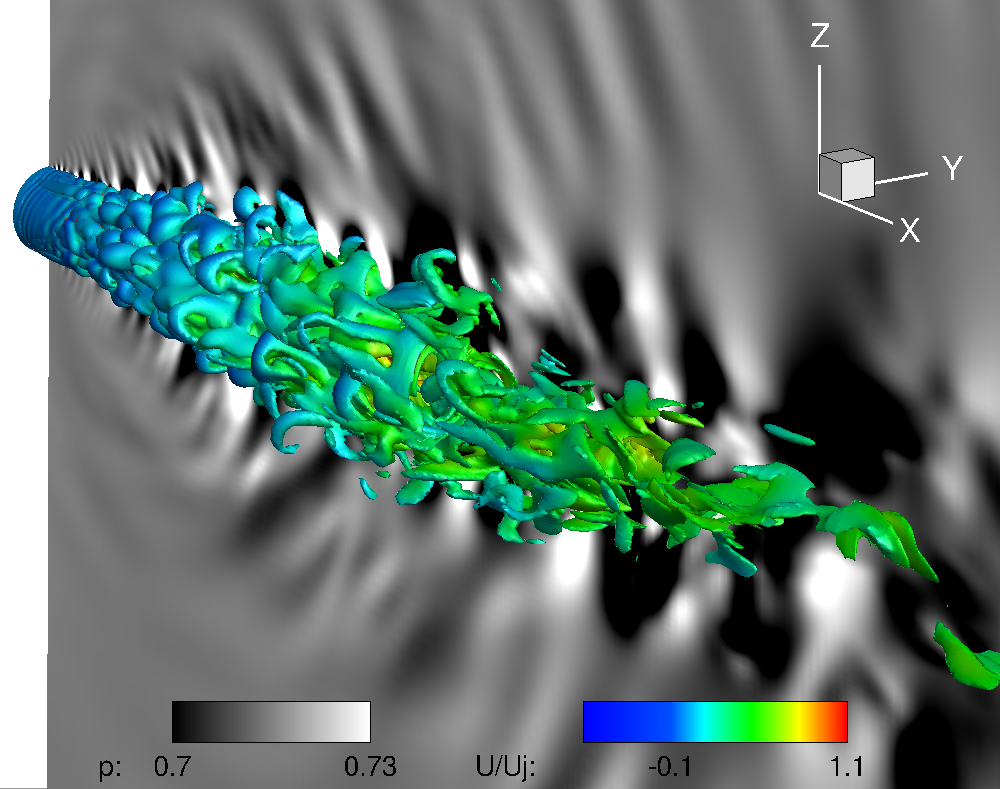}
    }
	\caption{Instantaneous lateral view of pressure contours, 
    in gray scale, superimposed by 3-D velocity magnitude 
    contours, in color.}
	\label{fig:3d-vel}
\end{figure}

\begin{figure}[htb!]
  \centering
  \subfigure[Lateral view of pressure and magniture of vorticity.]
    {\includegraphics[trim= 5mm 5mm 5mm 5mm, clip, width=0.45\textwidth]{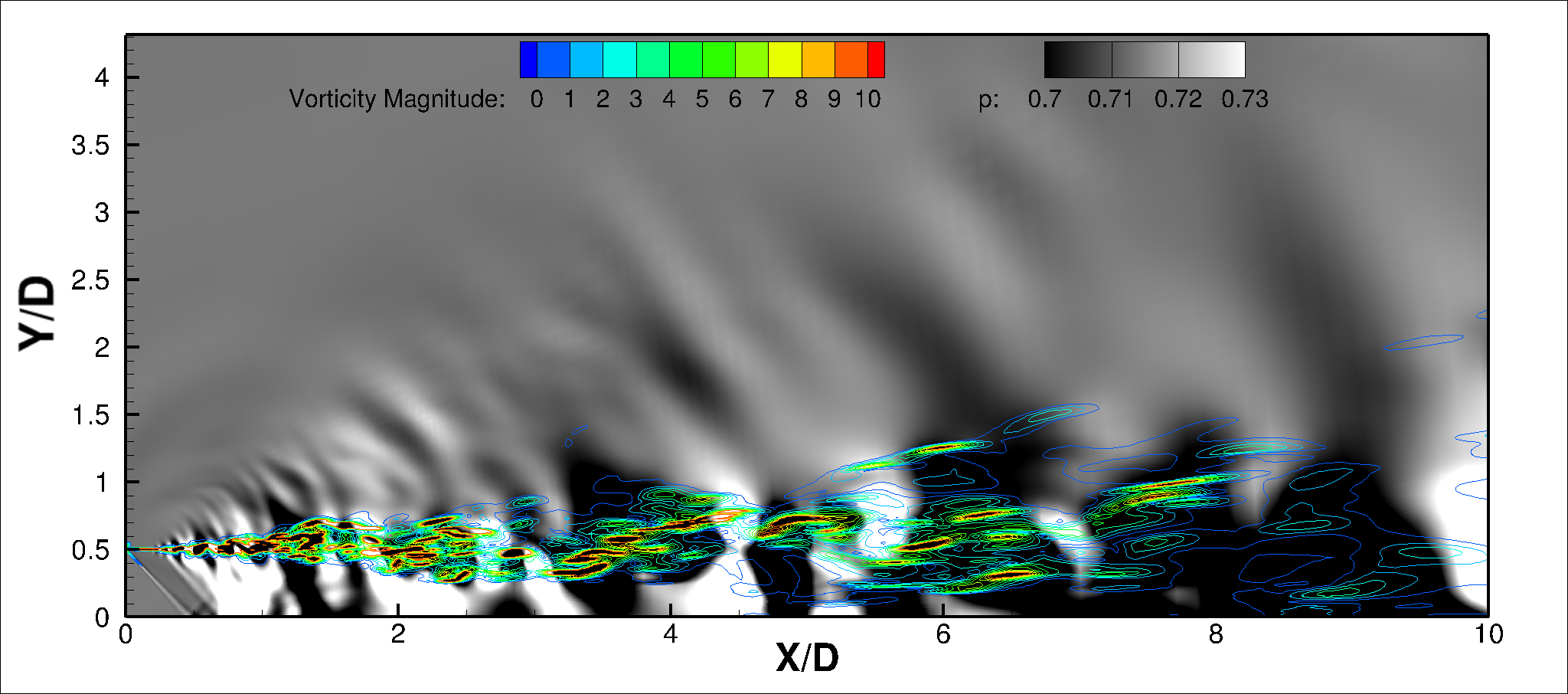}\label{fig:vort}}
	\subfigure[Detailed view of pressure and magnitude of 
    vorticity at the {jet exit of the nozzle}.]
    {\includegraphics[trim= 5mm 5mm 5mm 5mm, clip, width=0.45\textwidth]{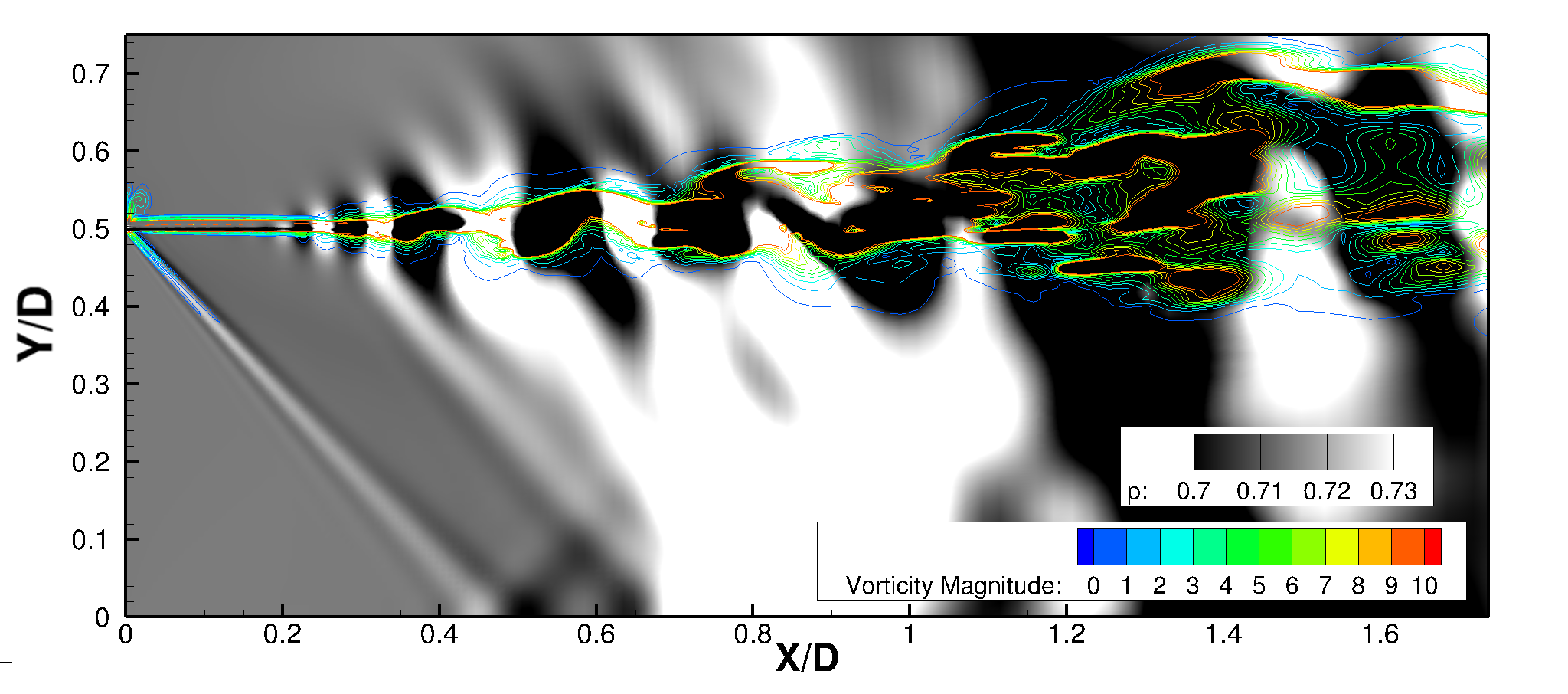}\label{fig:vort-zoom}}
	\caption{Lateral and detailed view of pressure contours, in 
    gray scale, superimposed by vorticity magnitude contours, 
    in color.}
	\label{fig:press-vort}
\end{figure}

Figures \ref{fig:3d-vel} and \ref{fig:press-vort} present a 
lateral view of an instantaneous visualization of the 
pressure contours, in gray scale, superimposed by 3-D 
velocity magnitude contours and vorticity magnitude 
contours respectively, in color, calculated by the LES tool 
discussed in the present paper. A detailed visualization of 
the entrance domain is shown in Fig.\ 
\ref{fig:vort-zoom}. The resolution of flow features 
obtained from the jet simulation is more evident in this 
detailed plot of the jet entrance. One can clearly notice 
the compression waves generated at the shear layer, and 
their reflections at the jet axis. Such resolution is 
important to observe details and behavior of such flow 
configuration in order to understand the acoustic phenomena 
which is present in supersonic jet flow configurations.



\section{Concluding Remarks}

The present work is concerned with an evaluation of the 
performance of a computational fluid dynamics tool for 
aerospace applications when using a national supercomputer 
from the Brazilian National Laboratory for Scientific 
Computing (LNCC). The cluster is named Santos 
Dumont and it is in the 193rd position of the top 500 list 
of November 2019.  
The numerical solver was developed by the authors to study 
supersonic jet flow configurations using a large eddy 
formulation. Results obtained with the code are used for 
aeroacoustic design applications. 
The simulations of such flow configurations 
are expensive and require continuous improvement of the 
numerical tool regarding efficient parallel computing. 
Therefore, weak and strong scalability studies of the 
solver are performed on the Santos Dumont supercomputer 
in order to evaluate if the numerical tool is capable of 
efficiently using thousands of processors in parallel. 
It should be noted that study is performed in the 
Teraflop partition of the Santos Dumont 
supercomputer which has 18,144 computational cores, 
396 Nvidia K40 graphical processing units and 108 Xeon 
Phi 7120 Intel accelerators. Moreover, the system section 
has a LINPACK performance of 456.8 TFlop/s and a theoretical 
peak performance of 657.5 TFlop/s.

Spatial discretization in the solver uses a second-order centered 
finite difference approach. Time integration is performed using a 
five-stage explicit Runge-Kutta scheme. The code is implemented 
using Fortran 90 standards coupled with message passing interface 
(MPI) for inter-partition communications. 
The jet flow-like geometry and flow condition were defined for the 
scalability study, and 13 different grid refinement levels were 
constructed. Furthermore, different partitioning configurations 
were used in order to evaluate the parallel code under different 
workloads. Grid sizes were varied from $370,000$ to approximately 
$1.0$ billion grid points. Calculations were performed for 1000 
time steps or 24 wall-clock hours of computation, whichever was 
reached first, using up to 3072 cores in parallel. The CPU 
time per iteration was averaged, when the simulation was finished, 
in order to calculate the speedup and scaling efficiency.
A preliminary strong scaling study is performed using 
a smaller computer to evaluate the effects from the recent 
implementations added to the LES solver using up to 400 
computational cores in parallel. Hence, more than 400 
simulations were performed for the scalability study of the 
parallel solver in the Santos Dumont supercomputer. 

The preliminary strong scalability study presents an increase of 
20\% on the efficiency curve when using a configuration of 50-million 
grid points along with up to 400 computational cores in 
parallel. The strong and weak scaling studies performed in the 
Santos Dumont computer
indicate that the parallel tool has a good strong scalability 
and it is able to speedup the time-to-solution when running on 
more than 3000 processing units with a strong scalability efficiency
of approximately 70\%. Super-linear strong scaling is also observed 
in the tests performed. Moreover, the largest mesh configuration 
addressed in the present effort has achieved a strong scalability 
which follows the ideal case very close, even when running on 2048 
computational cores. 

Five different workloads are considered in the present work in 
order to study the weak scaling of the solver. Tests are 
performed starting with $165 \times 10^{3}$ grid points per 
computing unit and moving up to $3.0 \times 10^{6}$ grid points 
per core. The load is doubled for each weak scalability test to 
the next one. The first two test cases present a significant 
decay on the weak scaling efficiency when increasing the number 
of processors. The performance for a fixed number of processors 
is improved for higher workloads. 
In the worst scenario evaluated in the present work, the test 
cases using 1.5 and 3.0 million points per core present a weak 
scalability efficiency of $\approx 70\%$. It is also important to 
remark that super-linear speedup events indicated in the strong 
scalability studies are related to the presence of efficiency 
peaks in the weak scaling curves. 

The results, obtained for the physically relevant test case, 
indicate that it is possible to achieve good results for 
supersonic jet flows using the present solver. The simulations 
performed to validate the numerical code are in good agreement 
with experimental and numerical references in the regions where 
the grid presents high resolution. Additionally, the present work 
indicates that the parallel implementation of the code is capable 
of handling high spatial resolution properly. Furthermore, the 
scalability study performed in the current paper can serve as a 
guide for future simulations using the same numerical 
tool in the Santos Dumont supercomputer. 
Nevertheless, the authors recognize the existence of improvement 
possibilities and will continue the development of the solver 
regarding the optimization of cache memory access and the 
implementation of vectorization capabilities.





\section*{Acknowledgments}
\label{acknowledgments}

The authors gratefully acknowledge the partial support for this 
research provided by Conselho Nacional de Desenvolvimento Cient\'{i}fico 
e Tecnol\'{o}gico, CNPq, under the Research Grants \# 309985/2013-7, 
400844/2014-1 and 443839/2014-0. The authors are also indebted to the 
partial financial support received from Funda\c{c}\~{a}o de Amparo 
\`{a} Pesquisa do Estado de S\~{a}o Paulo, FAPESP, under the Research 
Grants \# 2013/07375-0 and 2013/21535-0. The authors further acknowledge 
the National Laboratory for Scientific Computing, LNCC/MCTIC, for 
providing high-performance computing resources through the Santos Dumont 
supercomputer, which have been indispensable in order to obtain the 
research results reported in this paper.



\bibliographystyle{elsarticle-num}
\bibliography{main}


%
%
%


\section*{About the Authors}

{\bf Carlos Junqueira Junior} is a Research Engineer at 
\'{E}cole Nationale Sup\'{e}rieure d'Arts et M\'{e}tiers 
(ENSAM), in Paris. He graduated with the Engineering 
degree at the \'{E}cole Nationale Sup\'{e}rieure de 
l'\'{E}nergie l'Eau et l'Environnement (ENSE3) and also 
at the Universidade Estadual Paulista (UNESP). He holds 
the titles of Master of Science and Doctor of Science 
both achieved at Instituto Instituto Tecnol\'{o}gico de 
Aeron\'{a}utica. His research interests are in the area of 
computational fluid dynamics, high performance computing, 
and numerical methods.

{\bf Jo\~{a}o Luiz F. Azevedo} is a Senior Research 
Engineer in the Aerodynamics Division of Instituto de 
Aeron\'{a}utica e Espa\c{c}o, at S\~{a}o Jos\'{e} dos 
Campos, Brazil. His professional experience includes 
development and application of CFD codes for applied 
aerodynamic and aeroelastic analyzes of aerospace vehicles, 
aeroelastic clearance of launch vehicles, and aerodynamic 
CFD analyzes of wind tunnel models prior to testing. Areas 
of research interest include development of high-order, 
adaptive, unstructured grid CFD codes for realistically 
complex configurations, implementation of turbulence 
models, and development of cost effective techniques 
for coupling CFD solvers with aeroelastic analysis 
procedures.

{\bf Jairo Panetta} is a Professor of the Scientific Computing
Department at Instituto Tecnol\'{o}gico 
de Aeron\'{a}utica (ITA). He graduated as an engineer and 
received his Master of Science Degree, both at the Instituto 
Tecnol\'{o}gico de Aeron\'{a}utica. He defended his Ph.D. at 
Purdue University. His research interests are in the area of 
computer architecture, analogic processing and high performance 
computing.

{\bf William R. Wolf} is a Professor of the Faculdade 
de En\-ge\-nha\-ria Mec\^{a}nica at Universidade Estadual de 
Campinas (U\-NI\-CAMP). He holds a Mechanical Engineering 
degree obtained from the University of S\~{a}o Paulo and 
the title of Master of Science from Instituto 
Tecnol\'{o}gico de Aeron\'{a}utica. He defended his Ph.D. 
Thesis at Stanford University. His research interests are 
in the area of aeroacoustics, turbulent flows, aerodynamics, 
and computational fluid dynamics.

{\bf Sami Yamouni} is a Data Scientist at {Data{L}ab 
Serasa Experian. He has an Engineering degree from Institut 
Sup\'{e}rieur de M\'{e}canique de Paris and a Master of 
Science degree from the Universit\'{e} de Poitiers. 
Yamouni defended his Ph.D. Thesis at \'{E}cole 
Polytechnique. His research interests are in the area 
of dynamic mode decomposition, proper orthogonal 
decomposition, artificial intelligence and high performance 
computing.


\end{document}